\documentclass[reprint,amsmath,amssymb,aps,prr]{revtex4-1}
\usepackage{amsmath,amssymb,amsfonts,bm}
\usepackage{graphicx}
\usepackage{epstopdf}
\usepackage{dcolumn}
\usepackage{mathrsfs}
\usepackage{color}
\usepackage{multirow}
\usepackage{threeparttable}
\usepackage{appendix}

\usepackage[colorlinks=true,linkcolor=blue,citecolor=blue, urlcolor=blue,bookmarks=false]{hyperref}

\begin{document}

\title{Implementation of quantum stochastic walks for function approximation, two-dimensional data classification, and sequence classification}
\author{Lu-Ji Wang}
\affiliation{Institute for Brain Sciences and Kuang Yaming Honors School, Nanjing University, Nanjing 210023, China}
\affiliation{Department of Physics, Nanjing University, Nanjing 210093, China}
\author{Jia-Yi Lin}
\affiliation{Institute for Brain Sciences and Kuang Yaming Honors School, Nanjing University, Nanjing 210023, China}
\affiliation{Department of Physics, Nanjing University, Nanjing 210093, China}
\author{Shengjun Wu}
\affiliation{Institute for Brain Sciences and Kuang Yaming Honors School, Nanjing University, Nanjing 210023, China}
\affiliation{Department of Physics, Nanjing University, Nanjing 210093, China}

    \date{\today}
    \begin{abstract}
We study a quantum stochastic neural network (QSNN) based on quantum stochastic walks on a graph, and use gradient descent to update the network parameters.
We apply a toy model of QSNN with a few neurons to the problems of function approximation, two-dimensional data classification, and sequence classification.
A simple QSNN with five neurons is trained to determine whether a sequence of words is a sentence or not, and we find that a QSNN can reduce the number of training steps. A QSNN with 11 neurons shows a quantum advantage in improving the accuracy of recognizing new types of inputs like verses.
Moreover, with our toy model, we find the coherent QSNN is more robust against both label noise and device noise, compared with the decoherent QSNN.
These results show that quantum stochastic walks may be a useful resource to implement a quantum neural network.
	\end{abstract}
	
	\maketitle
	
	\section{Introduction}
        Propelled by rapid advances in computer hardware and artificial intelligence (AI) algorithms,
        neural networks (NN) \cite{lecun2015deep,jordan2015machine,goodfellow2016deep,nielsen2015neural} have found numerous applications in image processing \cite{lecun1998gradient, le2013building}, natural language processing \cite{hochreiter1997long, bengio2003neural,mikolov2011extensions,graves2013speech}, bioinformatics \cite{min2017deep}, etc.
        However, a bottleneck of classical computation may be right around the corner.
        Because it is too expensive to get a significant performance boost in the integration of chip components, and the classical physical rules do not apply when the scale is getting smaller and smaller.
        Compared with classical computation, quantum computation \cite{nielsen2002quantum,ladd2010quantum, arute2019quantum,zhong2020quantum} has displayed superiority in some specific problems such as factorization of large numbers \cite{shor1996fault}, search problems \cite{grover1997quantum}, and solving linear systems of equations \cite{harrow2009quantum}. More quantum algorithms and their efficient applications have been reviewed in Ref. \cite{montanaro2016quantum}.
        Furthermore, many studies \cite{biamonte2017quantum,dunjko2018machine} have combined quantum computation with machine learning to explore the potential of quantum AI.

        Some quantum generalizations of classical feedforward   \cite{situ2020quantum,narayanan2000quantum,da2016quantum,li2020quantum,zeng2019learning, farhi2018classification, mitarai2018quantum,dallaire2018quantum, killoran2019continuous, grant2018hierarchical, tacchino2019artificial, wan2017quantum,zhao2019building, he2021variational,beer2020training,steinbrecher2019quantum,cong2019quantum, dalla2020quantum} and feedback \cite{schuld2014quantum,amin2018quantum, rebentrost2018quantum,tang2019experimental,carleo2017solving,gao2017efficient} neural networks called quantum neural networks (QNNs) have been proposed.
        They have shown their abilities in quantum tasks, such as quantum automatic encoder
        \cite{wan2017quantum,steinbrecher2019quantum, bondarenko2020quantum}, quantum-state classification \cite{farhi2018classification,zhao2019building,cong2019quantum}, learning unknown unitaries \cite{he2021variational,beer2020training,steinbrecher2019quantum}, the simulation of many-body systems \cite{carleo2017solving, gao2017efficient}, and so on \cite{dalla2020quantum}.
        What's more, a number of explorations have shown the applications of the QNNs to learn from classical data \cite{killoran2019continuous,grant2018hierarchical,situ2020quantum,narayanan2000quantum,da2016quantum,li2020quantum,zeng2019learning, mitarai2018quantum,farhi2018classification,dallaire2018quantum,tacchino2019artificial, rebentrost2018quantum, amin2018quantum}.

        In addition to the QNNs used to complete fundamental tasks mentioned above, some quantum theories have been used to explain the inherent behavior in macro-world problems.
        For example, the probabilistic framework of quantum theory has been used to construct quantum language models \cite{zhang2018end, sordoni2013modeling, basile2017towards, zhang2019quantum}.
        These quantum language models are efficiently used in question answering \cite{zhang2018end}, ad-hoc retrieval \cite{sordoni2013modeling}, speech recognition \cite{basile2017towards}, and sentiment analysis \cite{zhang2019quantum}.
        These works inspire us to use QNNs to bridge the gap between the quantum computation and the language model, and search for advantages of the language models based on QNNs further.

        In this paper, we study a universal quantum stochastic neural network (QSNN) based on the quantum stochastic walks on a graph, and use gradient descent to optimize the evolution of the network.
        We show the QSNN's universal applicability and general application method by applying it to a series of classical tasks: function approximation, two-dimensional (2D) data classification, and sequence classification.
        Then, taking sentence recognition, a specific sequence classification task, as an example,
        we investigate the advantage of the QSNN by numerically comparing the training performance of three toy models with 5 neurons: the coherent QSNN, the decoherent QSNN, and a classical NN.
        Another comparison of these 3 kinds of networks with 11 neurons is implemented to explore the quantum advantage of the QSNN in recognizing verses.
        In the end, we evaluate how the QSNNs behave in the above comparative studies in the presence of label noise and device noise.

	\section{Quantum Stochastic Neural Network}\label{sec:model}
	
        \begin{figure*}[htp]
          \centering
          \includegraphics[width=18cm]{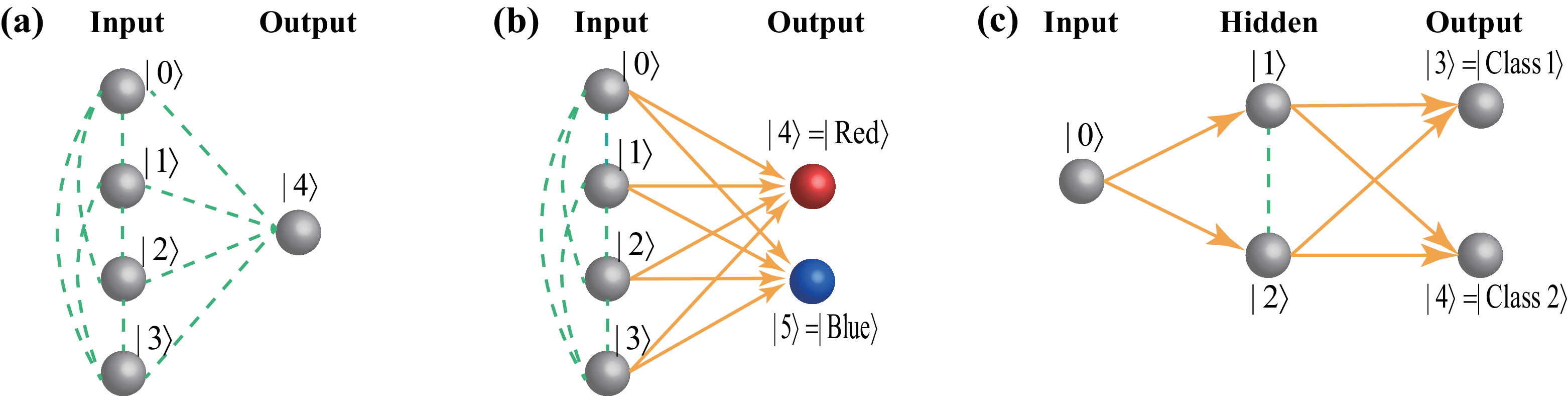}
          \caption{QSNN structures for three tasks. Neurons represent basis states $\{|i\rangle \}$ (for example, the first neuron is represented by the first basis state $|0\rangle=(10000)^{\intercal}$) of the Hilbert space and they are connected with each other by matrix elements of the Hamiltonian (green) or by Lindblad operators (orange).
          $N_{\text{in}}$ neurons are in the input layer, and the rest are in the hidden layer(s) and output layer.
          (a) The QSNN structure for approximation of functions.
          (b) The QSNN structure for classification of 2D data. Here $|\text{red}\rangle=|4\rangle=(000010)^{\intercal}$ and $|\text{blue}\rangle=|5\rangle=(000001)^{\intercal}$ represent the two categories that data will be divided into.
          (c) The QSNN structure for sequence classification with only 2 objects.
          The QSNN is initialized in the state $| 0\rangle$ (representing the single neuron in the input layer).
          The two neurons represented by $|1\rangle=(01000)^{\intercal}$ and $|2\rangle=(00100)^{\intercal}$ in the hidden layer correspond to the two different objects $w_1, w_2$.
          The output layer tells whether the sequence belongs to Class 1 $(|3\rangle =|\text{Class 1}\rangle)$ or Class 2 $(|4\rangle =|\text{Class 2}\rangle)$.}
          \label{fig:model}
        \end{figure*}

        Our quantum stochastic neural network (QSNN) is based on quantum stochastic walks (QSWs)\cite{whitfield2010quantum}, i.e. continuous-time quantum walks with decoherence, which have been used to construct other quantum networks in Ref. \cite{dalla2020quantum}.

        The state of a QSNN with $N$ neurons is represented by a density operator $\rho$ on an $N$-dimension Hilbert space $\mathcal{H}$, and $\{|i\rangle\}_{i=0}^{N-1}$ is an orthogonal basis of $\mathcal{H}$, with each basis state $|i\rangle$ corresponding to a neuron (vertex) of the QSNN.
        As shown in Fig.\,\ref{fig:model}, the QSNNs are structured in different ways to accomplish different tasks, but they generally consist of an input layer, several hidden layers (possibly a single hidden layer or non hidden layer), and an output layer.
        The state of the QSNN is initialized according to the particular task, and evolves according to the Gorini-Kossakowski-Lindblad-Sudarshan (GKLS) master equation \cite{kossakowski1972quantum, lindblad1976generators, gorini1976completely}
        \begin{equation}\label{eq:master equation}
        \frac{d\rho}{dt}=-i[H,\rho]+\sum_{k} \left(L_k\rho L_k^{\dagger}-\frac{1}{2}\{L^\dagger_k L_k,\rho\} \right).
        \end{equation}
        The first term on the right side determined by the Hamiltonian $H$ represents the coherent part of the dynamic (a factor $1/ \hbar$ is absorbed in the Hamiltonian), while the second term determined by the Lindblad operators $L_k$ represents the decoherent part, i.e., the coupling of the QSNN to its environment or a measuring device.
        In our approach, we use the Lindblad operators
        \begin{equation}\label{eq:channel}
        \begin{aligned}
        L_k \rightarrow L_{ij}=\gamma_{ij}|i\rangle\langle j |
        \end{aligned}
        \end{equation}
        to simulate the decoherent (typically one-way) transmission from the $j$th neuron to the $i$th neuron, and the Hamiltonian
        \begin{equation}\label{eq:hamiltonian}
        \begin{aligned}
        H=\sum_{ij}h_{ij}|i\rangle\langle j|
        \end{aligned}
        \end{equation}
        to characterize the coherent transmission between neurons, like a feedback process of Hopfield networks but in a completely quantum way.
        The coefficients $\gamma_{ij}$ that characterize the dissipation rates from neuron $|j\rangle$ to neuron $|i\rangle$ are real numbers in general, while the coefficients $h_{ij}$ are complex numbers with the requirement $h_{ij}=h_{ji}^{*}$ to ensure the Hermiticity of the Hamiltonian. For the tasks of our interest, real coefficients $h_{ij} \in\mathbb{R}$ are sufficient. Therefore, the Hamiltonian is written as
        \begin{equation}\label{eq:hamiltonian1}
        \begin{aligned}
        H=\sum_{ij}h_{ij}|i\rangle\langle j|=\sum_{i<j}h_{ij}(|i\rangle\langle j|+|j\rangle\langle i|)
        \end{aligned}
        \end{equation}
        when there is no transmission from a neuron to itself.

        We group all the real parameters in the Hamiltonian as a single real vector $\vec{h}=(h_1,h_2,\cdots,h_m)\in\mathbb{R}^m$ and the real parameters in the Lindblad operators as another real vector $\vec{\gamma}=(\gamma_1,\gamma_2,\cdots,\gamma_n)\in\mathbb{R}^n$, where $m$ ($n$) is the number of independent
        real parameters in the Hamiltonian (Lindblad operators).

        For the convenience of numerical manipulation, we can rewrite
        Eq.\,(\ref{eq:master equation}), according to the Choi-Jamiolkowski isomorphism, as
        \begin{equation}\label{eq:cj master equation}
        \frac{d|\rho\rangle}{dt}=\mathcal{L}|\rho\rangle
        \end{equation}
        where $|\rho \rangle =\sum_{ij} \rho_{ij} |i\rangle |j \rangle$ is the ket in $\mathcal{H}\otimes \mathcal{H}$ corresponding to the density matrix $\rho= \sum_{ij} \rho_{ij} |i\rangle \langle j |$, and
        the operator $\mathcal{L}$ corresponding to the Liouvillian superoperator is given by
        \begin{equation}
        \begin{aligned}
        \mathcal{L}(\vec{h}, \vec{\gamma})=&-i(H\otimes I-I\otimes H^T) \\
        &+ \sum_k ( L_k\otimes L_k^{*}-\frac{1}{2}L_k^\dagger L_k\otimes I-\frac{1}{2}I\otimes L_k^TL_k^* ).
        \end{aligned} \label{Loperator}
        \end{equation}
        Here, $A^*$, $A^T$, and $A^\dagger$ denote, respectively, the complex conjugate, the transpose, and the adjoint of $A$.

        After encoding the classical data into the initial state of the QSNN, we optimize the parameters (coefficients $h_k, \gamma_k$) governing the evolution process with a learning algorithm based on gradient descent.
        In order to explore the universality of the QSNN, we use it to approximate several functions in the following subsection A. After that, we also apply the QSNN to two more complicated tasks, namely the 2-D data classification in subsection B and the sequence classification in subsection C.

        \subsection{Approximation of functions}
        A key step for a QSNN to process classical tasks is to encode the classical input data in the initial state of the QSNN, which is a significant research topic itself.
        Generally, it is natural to encode an unnormalized $n$-dimensional real vector $\vec{x}=(x_0, x_1, \cdots, x_{n-1})\in\mathbb{R}^n$ into the initial state of $N_{\text{in}}=n+1$ input neurons (i.e., basis states).
        In order to introduce non-linearity, we choose to introduce higher-order terms of classical input data with the encoding as follows.
        For the $s$th training sample from the training set, represented by an $n$-dimensional real vector $\vec{x}_s= (x_0, x_1, \cdots, x_{n-1})_s $, we choose to initialize the QSNN (shown in Fig.\,\ref{fig:model}(a)) in the state
        \begin{equation}\label{eq:encode}
        |\psi_s\rangle=\frac{1}{C_s}\sum_{i=0}^{K-1}\sum_{d=0}^{n-1}(x_{d})_s^i|d K +i\rangle,
        \end{equation}
        where we choose the integer $K=\frac{N_{\text{in}}}{n}$ to be greater than $1$ to ensure that the complete information of the input is encoded, and $C_s$ is the normalization factor.
        A larger $K$ introduces more higher order terms of the classical input data so that the QSNN can fit more complex mappings (functions) in general.
        For the $s$th sample, the initial density operator of the QSNN is hence $\rho_{\text{in}}^s=|\psi_s\rangle \langle\psi_s|$, and its corresponding vector with respect to the Choi-Jamiolkowski isomorphism is $| \rho_{\text{in}}^s\rangle =|\psi_s\rangle \otimes |\psi_s \rangle$.

        To start with, we consider the simplest case where the input $x$ is a single real number ($n=1$), and the QSNN is trained to approximate a real-valued analytic function $f(x)$. This is a common task to demonstrate the universality of classical neural networks \cite{cybenko1989approximation, hornik1989multilayer} and quantum learning models \cite{mitarai2018quantum}.

        To approximate a function $f(x)$, we need to know some samples $x_s$ paired with their values $f(x_s)$ under the function $f$.
        These samples constitute a training set $\{(x_s, f(x_s))\in\mathbb{R}^2\}_{s=1}^M$, where $M$ is the number of pairs in the training set.
        For each sample, we encode the input data $x_s$ in the initial state of the QSNN according to Eq.\,(\ref{eq:encode}), and let the QSNN evolve according to Eq.\,(\ref{eq:master equation}) for a duration $T$.
        As shown in Fig.\,\ref{fig:model}(a), the $N$ neurons are only coherently connected with each other via the Hamiltonian for this function approximation task.
        So only the parameters $\vec{h}$ for the Hamiltonian can be non-zero and will be updated to optimize the evolution process.
        All the parameters $\vec{\gamma}$ for the Lindblad operators are set to be zero for this task.
        The state of the QSNN after the above evolution $\Lambda(\vec{h})$ can be given as
        \begin{equation}
        |\rho_{\text{out}}^s\rangle=\Lambda(\vec{h})|\rho_{\text{in}}^s\rangle=e^{\mathcal{L}(\vec{h})T}|\rho_{\text{in}}^s\rangle
        \end{equation}
        by solving Eq.\,(\ref{eq:cj master equation}), where the operator $\mathcal{L}(\vec{h})$ is given by Eq.\,(\ref{Loperator}) with $H=\sum_{i,j=0;\; i<j}^{N-1}h_{ij}(|i\rangle\langle j|+|j\rangle\langle i|)$, and $L_k =0$.

        \begin{figure}[pt]
          \centering
          \includegraphics[width=9.5cm]{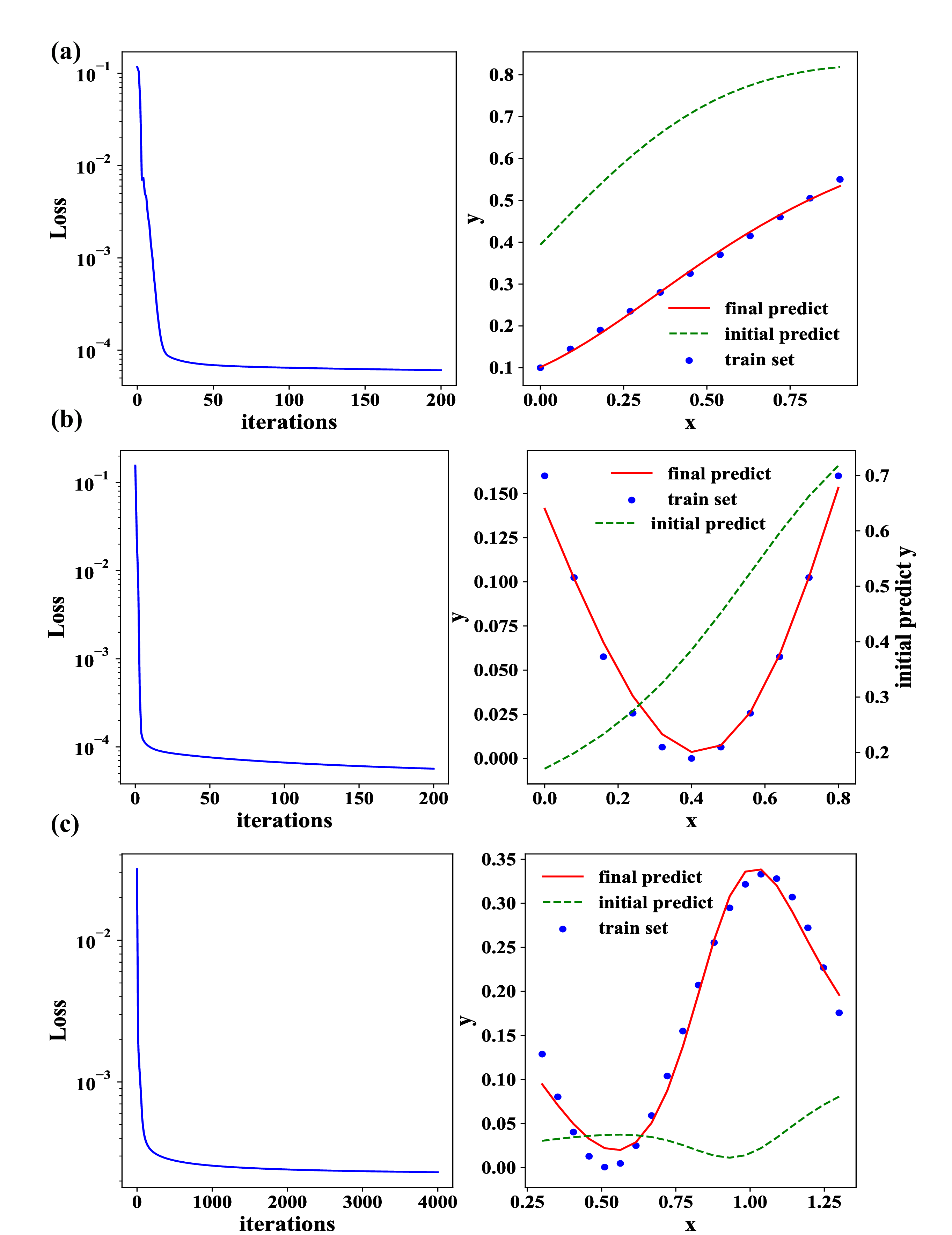}
          \caption{Performance of the QSNN to approximate the functions (a) $f(x)=0.5x+0.1$, (b) $f(x)=(x-0.4)^2$, and (c) $f(x)=(1+\cos{6x})/6$.
          The number of neurons in the input layer of QSNNs is $N_{\text{in}}=2, 4, 10$ in the simulation of (a), (b), and (c), respectively. The blue points in the right column of the figure are the training data.
          The green dotted lines are given by the untrained QSNNs with random parameters.
          After training, the output of QSNNs is closed to the desired output as shown in the red lines.}
        \label{fig:fit}
        \end{figure}

        After the evolution, a projection onto the output neuron is performed by measuring $\Omega=|N-1\rangle\langle N-1|$ on the final state $\rho_{\text{out}}^s$ to give the final output $\text{Tr}(\Omega\rho_{\text{out}}^s)$ of the QSNN for the $s$th sample from the training set.
        In order to minimize the distance between the output $\text{Tr}(\Omega\rho_{\text{out}}^s)$ of the QSNN and the ideal target output $f(x_s)$, we define the loss function as
        \begin{equation}
        \mbox{Loss}=\frac{1}{M}\sum_{s=1}^M\left[\text{Tr}(\Omega\rho_{\text{out}}^s)-f(x_s)\right]^2  .
        \end{equation}
       We train the QSNN by updating the parameters of $\vec{h}$ with a learning algorithm based on gradient descent.

        The gradient of the loss function with respect to any parameter $\theta$ can be written as
        \begin{equation}\label{eq:gradient1}
        \frac{\partial \mbox{Loss}}{\partial \theta}= \frac{1}{M}\sum_{s=1}^M2\left[\text{Tr}(\Omega\rho_{\text{out}}^s)-f(x_s) \right]\text{Tr}(\Omega\frac{\partial\rho_{\text{out}}^s}{\partial \theta}).
        \end{equation}
        Then the parameter can be updated according to
        \begin{equation}\label{eq:update}
        \theta^\prime=\theta-\eta\frac{\partial \mbox{Loss}}{\partial\theta} ,
        \end{equation}
        where $\eta$ is an adjustable positive parameter called the learning rate.
        A detailed calculation method of the gradients $\frac{\partial\rho_{\text{out}}^s}{\partial \theta}$ in Eq.\,(\ref{eq:gradient1}) is provided in Appendix \ref{appendix:learning algorithm}.
        After a handful of epoches, the loss function will approach a local minimum.

        As shown in Fig.\,\ref{fig:fit}, our QSNN can approximate simple analytic functions. The more complex the function is, the more neurons in the input layer are required. In the simulation, the QSNNs are initialized with random parameters $\vec{h}$ ($h_k\in[0, 1]$ uniformly) and the dimensionless time duration $T$ of the evolution is set to be $1$.

\subsection{Classification of 2D data}\label{sec:classification of 2D data}

\begin{figure}
  \centering
  \includegraphics[width=9cm]{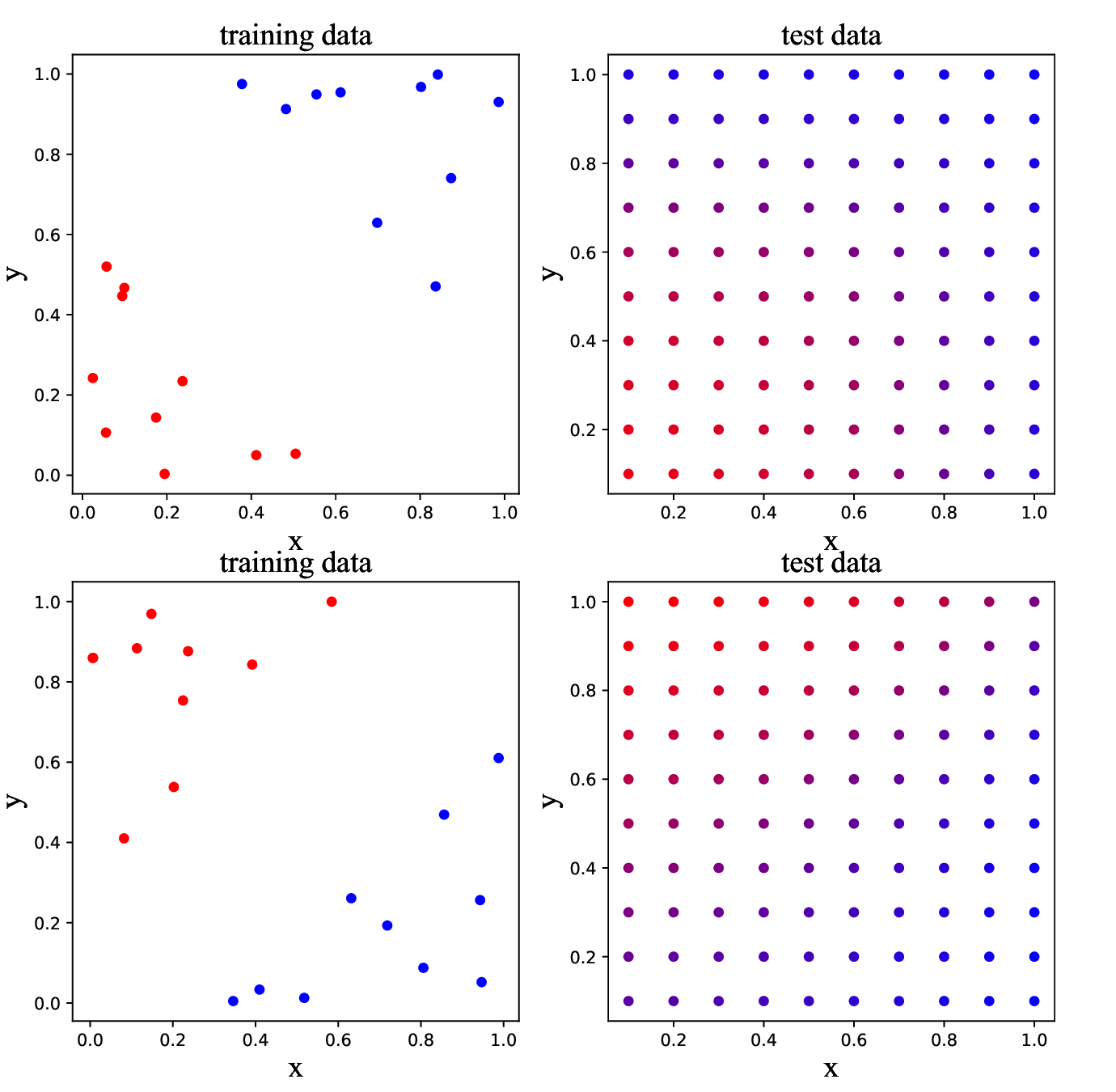}\\
  \caption{Classification of 2D data via a QSNN. The data are a set of 2-dimensional points, each point will be classified into one of two categories (red or blue). The left two images, with each point already labeled red or blue, represent two different training sets. A QSNN of Fig.\,\ref{fig:model}(b) with $N_{in}=12$ is trained with the training data from one of the left images. We consider a set of equally distributed 2D points as test data, and the upper (lower) right image contains the test data classified by the QSNN that is trained with the training data in the upper (lower) left image.}
\label{fig:2d classification}
\end{figure}

        The classification task is one of the most active research topics of neural networks in both classical \cite{zhang2000neural} and quantum \cite{farhi2018classification,mitarai2018quantum} domains.
        We use the QSNN to accomplish two classification tasks in this and the next subsection.
        We classify 2-dimensional (2D) classical data in this subsection. The training set is $\{\vec{r}_s, l_s)\}_{s=1}^M$. Each input
        $\vec{r}_s=(x_s, y_s)$ can be considered as a point on a plane.
        The points belong to one of the two categories as shown in the left column of Fig.\,\ref{fig:2d classification}.
        The label $l_s\in\{\text{red}, \text{blue}\}$ indicates which category the point $\vec{r}_s$ belongs to.
        We encode the input $\vec{r}_s$ into the initial state $\rho_{\text{in}}^s$ of the QSNN shown in Fig.\,\ref{fig:model}(b) according to Eq.\,(\ref{eq:encode}).
        2 neurons in the output layer of the QSNN represent 2 possible output labels (red, blue).
        We use one-way Lindblad operators to connect the input layer to the output layer so that probability is to be accumulated in the output layer.
        In addition, neurons in the input layer are also coherently connected by the Hamiltonian.
        The subsequent evolution of the QSNN is divided into the following stages:

        (1) \emph{Unitary evolution} $\Lambda^U(\vec{h})$. It is a coherent evolution for a duration $T^{U}$. The state of the QSNN after this coherent evolution is
        \begin{equation}\label{eq:unitary evolution}
       |\rho_U\rangle=\Lambda^U(\vec{h})|\rho_{\text{in}}\rangle=e^{\mathcal{L}_H(\vec{h}) T^U}|\rho_{\text{in}}\rangle,
        \end{equation}
        where $\mathcal{L}_H(\vec{h})$ is given by Eq.\,(\ref{Loperator}) with the Hamiltonian being $H=\sum_{i, j=0; i<j}^{N_{\text{in}}-1}h_{ij}(|i\rangle\langle j|+|j\rangle\langle i|)$ and Lindblad operators all being zero.

        (2) \emph{Dissipative evolution} $\Lambda^D(\vec{\gamma})$. It is a completely dissipative process for a duration $T^{D}$ with the Hamiltonian being zero and only Lindblad operators $\{L_{ij}=\gamma_{ij}|i\rangle\langle j|\}$ that connect from the input layer to the output layer being in effect. The state of the QSNN after the second stage is
        \begin{equation}\label{eq:output evolution}
        \begin{aligned}
        |\rho_{\text{out}}\rangle&=\Lambda^D(\vec{\gamma})|\rho_U\rangle=\Lambda^D(\vec{\gamma})\Lambda^U(\vec{h})|\rho_{\text{in}}\rangle\\
        &=e^{\mathcal{L}_D(\vec{\gamma}) T^D}e^{\mathcal{L}_H(\vec{h}) T^U}|\rho_{\text{in}}\rangle .
        \end{aligned}
        \end{equation}

         (3) \emph{Measurement}. After these two stages of evolution, we perform a measurement of $\Omega^s=|l_s\rangle\langle l_s|$ on the final state $\rho_{\text{out}}^s$ for the $s$th sample and define the loss function for a batch of $M$ samples as
        \begin{equation}
        \mbox{Loss}=1-\frac{1}{M}\sum_{s=1}^M\text{Tr}(\Omega^s\rho_{\text{out}}^s).
        \label{defloss}
        \end{equation}

        To minimize this loss,
        the parameters $\vec{h}$ and $\vec{\gamma}$ should be updated according to Eq.\,(\ref{eq:update}) in the direction opposite the gradient
        \begin{equation}\label{eq:gradient}
        \frac{\partial \mbox{Loss}}{\partial \theta}=-\frac{1}{M}\sum_{s=1}^M\text{Tr}(\Omega^s\frac{\partial\rho_{\text{out}}^s}{\partial \theta}),
        \end{equation}
        where $\theta$ represents any parameter $h_k$ or $\gamma_k$ to be updated.
        More details about calculating this gradient are provided in Appendix \ref{appendix:learning algorithm}.

        The results of applying our QSNN to the classification tasks are shown in Fig.\,\ref{fig:2d classification}.
        Each left image of Fig.\,\ref{fig:2d classification}, representing a training set,  contains a set of points labeled either red or blue in advance. Each right image, representing a test set, contains a set of 2D points that are colored according to the output of the QSNN trained with the data in the corresponding left image. Our QSNN with $14$ neurons ($12$ in the input layer and $2$ in the output layer) works well for the 2D classification task.
        All the nonzero $h_{ij}$ are initialized as $0.1$ for the unitary evolution, all the dissipation rates $\gamma_{ij}$ between neurons of the two different layers are also initialized as $0.1$ for the dissipative evolution, and $T^U=T^D=1$.

        \subsection{Sequence classification}\label{subsec:sentence recogition}
        A sequence is an important kind of data that is composed of a number of objects whose order does matter.
        There are many kinds of sequences that impact our lives, such as word sequences, signal sequences, and gene sequences.
        The different order of the objects causes a sequence to have different characteristics and to belong to different categories.

        To deal with sequence classification, we introduce a QSNN with three layers of neurons.
        The first layer contains a single neuron (corresponding to the state $|0\rangle$) and the network is initialized in this local state ($\rho_{0} = |0\rangle\langle 0|$).
        The second layer contains a certain number of neurons with each neuron corresponding to an object possibly appearing in a sequence we need to process.
        The single neuron in the first layer is connected to each neuron in the second layer by a one-way Lindblad operator. These Lindblad operators are turned on one after one temporally in the same order as the appearance of the objects in the sequence.
        The neurons in the second layer are also coherently connected with each other via the Hamiltonian.
        The final layer, i.e. the output layer, contains a certain number of neurons required to express the answer. Neurons from adjacent layers are connected by a one-way Lindblad operator.
        The time evolution of the QSNN is described by Eq.\,(\ref{eq:master equation}).

        For example, if the sequences to be classified contain only two different objects, the QSNN can be constructed by 5 neurons as shown in Fig.\,\ref{fig:model}(c).
        In this case, the classical training set is denoted as $S=\{(e_s, l_s)\}_{s=1}^2$, where the $s$th input (a sequence of the two objects) $e_s=(w_i, w_j)_{i\neq j\in\{1,2\}}$ is labeled
        $l_s\in\{\text{Class 1}, \text{Class 2}\}$.
        The QSNN processes a sample $(e_1=(w_1,w_2), l_1)$ with the following steps.

        (1) \emph{Input}: At time $t=0$, the network is initialized in the state $\rho_0=|0\rangle\langle0|$ and the channel controlled by $L_1^{\text{in}}=\gamma|1\rangle\langle0|$ is turned on. At $t= T^{\text{in}} /2 $, the channel controlled by $L_2^{\text{in}}=\gamma|2\rangle\langle0|$ is turned on. After this input process, the state of the network becomes
        \begin{equation}\label{eq:rho_in}
        \rho_{\text{in}}=\omega|0\rangle\langle0|+\alpha|1\rangle\langle1|+\beta|2\rangle\langle2|
        \end{equation}
        at $t=T^{\text{in}}$, where $\omega,\alpha,\beta$ are determined by $\gamma$ and $T^{\text{in}}$ (details in Appendix \ref{appendix:expression}).

        A more complex QSNN can contain more neurons that encode more different objects in the hidden layer. The encoding method above preserves the network's expandability and transferability.
        That is to say, it is possible to train a QSNN, which has been trained with a small data set before, with a larger one that consists of more objects. This can be done by introducing new neurons in the hidden layer of the trained network to encode new objects.

        (2) \emph{Evolution}: After the \emph{Input} stage of encoding, the QSNN evolves in two subsequent stages, i.e., the \emph{Unitary evolution} stage and the \emph{Dissipative evolution} stage, similar to the evolution stages in the classification task described in Sec. \ref{sec:classification of 2D data}. The final state after the evolution stages is given by Eq.\,(\ref{eq:output evolution}). However, the operator $\mathcal{L}_H$ governing the \emph{Unitary evolution} is determined by the Hamiltonian $H=h(|1\rangle\langle2|+|2\rangle\langle1|)$ here and the operator $\mathcal{L}_D$ governing the \emph{Dissipative evolution} stage is determined by the Lindblad operators $\{L_{ij}^{\text{out}}=\gamma_{ij}|i\rangle\langle j|\}$ with non-vanishing $\gamma_{ij}$ only when $i \in \{3,4 \} $ and $ j \in \{1,2 \}$.
        For classifying sequences with more than two objects, we deal with a hidden layer with more than two neurons, and we have a Hamiltonian with a nonzero matrix element connecting each pair of neurons in the same layer, as well as a non-vanishing Lindblad operator connecting each pair of neurons from two adjacent layers.

        (3) \emph{Measurement}: A projective measurement onto the output layer gives Class 1 with a probability $p_{\text{Class 1}} =\langle \text{Class 1} | \rho_{\text{out}} | \text{Class 1} \rangle$, Class 2 with a probability $p_{\text{Class 2}} =\langle \text{Class 2} | \rho_{\text{out}} | \text{Class 2} \rangle$, as well as an undetermined result with probability $1-p_{\text{Class 1}} -p_{\text{Class 2}}$ which is negligible for a sufficiently large $T^{\text{in}}$ and $T^D$.

        We train the QSNN to classify sequences by updating the parameters ($\vec{h}$ and $\vec{\gamma}$) governing the evolution with a learning algorithm based on gradient descent. Similar to Eq. (\ref{defloss}), the loss function is defined as
        \begin{equation}\label{eq:loss function}
        \mbox{Loss}=1-\frac{1}{M}\sum_{s=1}^M\text{Tr}(|l_s\rangle\langle l_s|\rho_{\text{out}}^s),
        \end{equation}
        where $M$ is the number of training samples.
        We try to minimize the above loss so that the QSNN can classify the sequences correctly with a higher probability.
        The gradient of the loss function is calculated according to Eq.\,(\ref{eq:gradient}).

        To investigate the ability of the QSNN in sequence classification, we demonstrate in Appendix \ref{appendix:expression} that the simple QSNN in Fig.\,\ref{fig:model}(c) (even with a vanishing Hamiltonian) can be trained to correctly classify any input sequence with a high probability, by showing that an optimal solution for the dissipation rates $\vec{\gamma}$ exists for any possible training set.

        We show the performance of the QSNN in sequences classification in the next section, taking the sentence recognition task as an example.

    \section{Advantages of quantum stochastic neural networks}\label{sec:quantum advantages}

        In this section, we investigate the performance of QSNN in sentence recognition.

        According to whether the Hamiltonian in Eq.\,(\ref{eq:master equation}) vanishes, we divide our QSNN into two categories: the coherent QSNN with both coherent evolution and dissipation evolution, and the decoherent QSNN with a zero Hamiltonian and only a dissipation evolution governed by the Lindblad operators.
        Since the classical NN has been relatively well developed, we don't expect our QSNN to completely surpass classical NN for all tasks.
        However, we like to understand the quantum advantage of QSNN.
        In our simulation, we compare the training and test performances of three types of neural networks, namely, the coherent QSNN, the decoherent QSNN, and an analogous classical NN.

        We use the sentence recognition task as an example to investigate the performance of the QSNN in handling a specific sequence classification task.
        Natural Language Processing (NLP) is an important application of classical AI.
        Sentence recognition, a fundamental NLP task, is to train the machine to identify whether a word sequence forms a meaningful sentence or not.
        So, the $s$th training data $e_s$ here is a sequence of words labeled $l_s\in\{{\text{Yes}, \text{No}}\}$ according to whether the sequence is a sentence or not.
        Here, we only consider the order and the collocations of words, not the different derived forms of words.
        To simplify our task and reduce the model size, the stop words such as prepositions and articles are also excluded from our consideration.
        Hence, we will lemmatize all the words in a sentence to their normal form and delete all stop words, before we process that sentence with the QSNN.

        \subsection{Training performance}\label{subsec:accelerate}

       In order to minimize the loss function (i.e., maximize the success probability of correct classification), the parameters need to be updated several times in the training process.
       We always pursue a fewer number of required training steps after which the loss function decreases to its local minimum.
       In this subsection, we demonstrate the advantage of the QSNN from the perspective of the number of training steps.

       We consider the 5-neuron model shown in Fig.\,\ref{fig:model}(c) with the training set $S=\{(e_1=(w_1, w_2), \text{Yes}),(e_2=(w_2, w_1), \text{No})\}$.
       We train both the coherent and decoherent QSNN from 1000 random initialization samples $\vec{\gamma}\in\mathbb{R}^n\ (\gamma_i\in[-1, 1]\ \text{uniformly})$, and average over the samples.

        \begin{figure}[h]
          \centering
          \includegraphics[width=8.5cm]{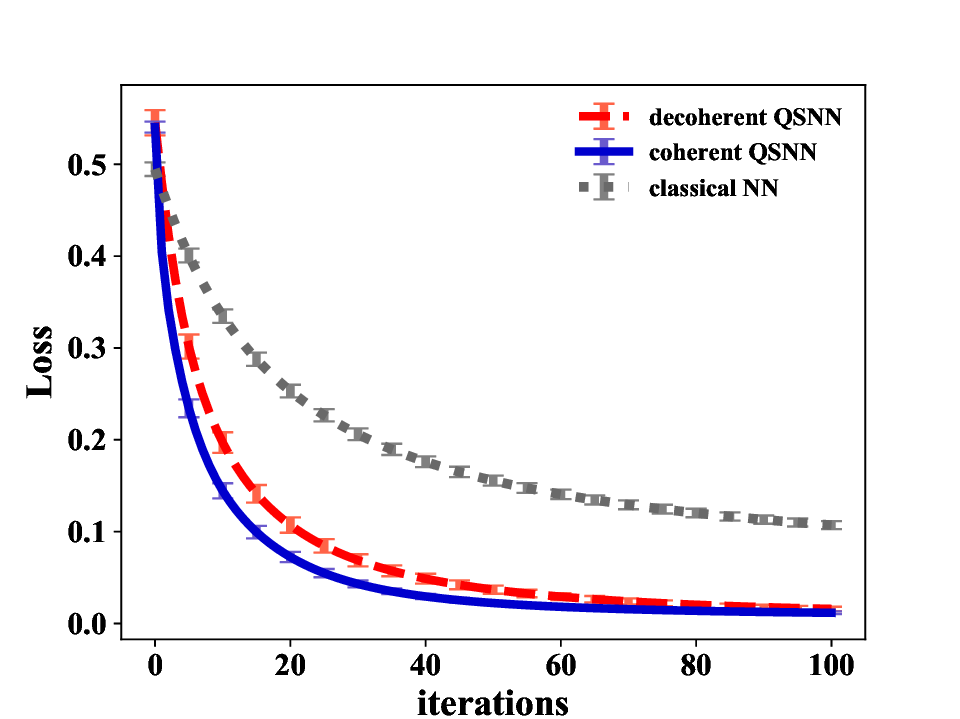}\\
          \caption{A plot of the loss function (the mean of 1000 initialization samples) of the networks with respect to the number of iterations used in the training procedure.
          Three curves in the figure respectively represent the results of three networks, i.e., the classical NN (grey), decoherent QSNN (red), and coherent QSNN with initial $h=0.1$ (blue).
          The coherent QSNN has a mean Loss going down faster than the decoherent QSNN, which is again faster than the classical NN.
          Namely, the coherent QSNN completes learning with a minimum number of training steps in sentence recognition.
          The error bars are drawn as 95\% confidence intervals. More simulation details are given in Appendix \ref{appendix:details about accelerate}.}
          \label{fig:accelerate fitting}
        \end{figure}

        The classical NN for comparison here is a typical feed-forward NN with 2 input neurons and 2 output neurons.
        For the $s$th input sample $e_s$, the input data of the classical NN is determined by the quantum state $\rho_{\text{in}}$ described as Eq. (\ref{eq:rho_in}).
        Namely, the data as the input of the classical NN is the probability distribution $(\alpha_s, \beta_s)$ over the neurons $|1\rangle$ and $|2\rangle$ of the QSNN after the \emph{Input} stage.
        Then, the classical NN governed by the weight $\vec{w}$ and bias $\vec{b}$ evolves.
        And we use the standard softmax function as the activation function to obtain a normalized output $\vec{y_s}(\vec{w}, \vec{b}) = (y_{\text{No}}, y_{\text{Yes}})$ of the classical NN.
        The output gives the probabilities for the input $e_s$ to be identified as a sentence ($y_{\text{Yes}}$) and not ($y_{\text{No}}$).
        The loss function of the classical NN for a batch of $M$ samples is defined as
        $$\mbox{Loss}_C=1-\frac{1}{M}\sum_{s=1}^M\vec{l_s}\cdot\vec{y_s},$$
        where $\vec{l_s}\in\{\text{No}=(1,0), \text{Yes}=(0, 1)\}$ is a classical label.

        The training result of the coherent QSNN, decoherent QSNN and the classical NN is shown in Fig.\,\ref{fig:accelerate fitting}.
        It can be found that the loss function of the QSNNs, both coherent and decoherent, converge faster than the classical NN.
        Then, we compare the training performances between the coherent and decoherent QSNN. Fig.\,\ref{fig:accelerate fitting} shows that the mean Loss always goes down faster when the model contains coherence.
        In summary, when the learning rates and the sizes of the three networks are the same, the number of training steps required for the two QSNNs with 5 neurons to achieve the same level of performance is less than the classical NN, and the coherent QSNN needs the least training steps.
        So, we infer that such a small size of the QSNN, especially the coherent one, can reduce the number of required training steps in the sentence recognition task.
        The result of this simulation will remain the same when the data of word sequences are replaced with other kinds of sequences.
        In order to exclude the influence of the number of parameters on the result, we have compared the coherent QSNN with another decoherent QSNN with the same number of parameters.
        We find that the coherent QSNN always performs better, regardless of whether the number of parameters of the decoherent QSNN is increased to the same as that of the coherent one.
        A more detailed discussion about the number of parameters is given in Appendix \ref{appendix:number of parameters}.

        With the above results, we believe that the quantum advantage of QSNN in training speed has some physical reasons, although it cannot be quantitatively characterized due to the difficulty of describing the complexity of neural network training \cite{livni2014computational}.
        Although our QSNN is based on the stochastic quantum walk,
        the speedup in training is different from the commonly known quadratic advantage in probability spread speed or mixing time of a quantum walk,
        as the latter focuses on the time required to perform a certain task with a fixed evolution,
        while the quantum advantage we discuss here is about the speed in updating the parameters to find the optimal evolution.
        Instead, the quantum advantage of QSNN in training speedup is more likely due to the same mechanism
        of the quantum advantage in shadow tomography \cite{aaronson2019shadow, kunjummen2021shadow, huang2021information}.
        The method of shadow tomography can reduce the number of experiments to learn a quantum evolution
        \cite{huang2021information, huang2021provably, huang2021quantum}, while our QSNN is also trained to learn a certain quantum evolution for a specific task. There could be a deeper relation between shadow tomography and quantum neural networks.
        Despite the differences between our work and the works on shadow tomography,
        the quantum advantages in various forms in the presence of quantum coherence may all be related in some intrinsic way,
        which itself deserves a deeper investigation.

        We also discuss the vanishing gradient problem \cite{glorot2010understanding,shalev2017failures,mcclean2018barren} when the QSNN is used to learn larger data sets in practical applications in Appendix \ref{appendix:vanishing gradient}.
        Because the initial value of the coherent strength $h$ has a certain influence on the training performance of the coherent QSNN, we provide a strategy to choose an appropriate initial value of $h$ rather than a random one.
        Details about the strategy are given in Appendix \ref{appendix:details about accelerate}.


         \subsection{Verse recognition}\label{sub:verse-recognition}
          A verse is a special kind of sentence with a complex and varied syntax, which makes it challenging to recognize verses by machines.
          Here, we give some evidence that the QSNN can improve the accuracy of verse recognition.

        \begin{figure}[h]
          \centering
          \includegraphics[width=8.5cm]{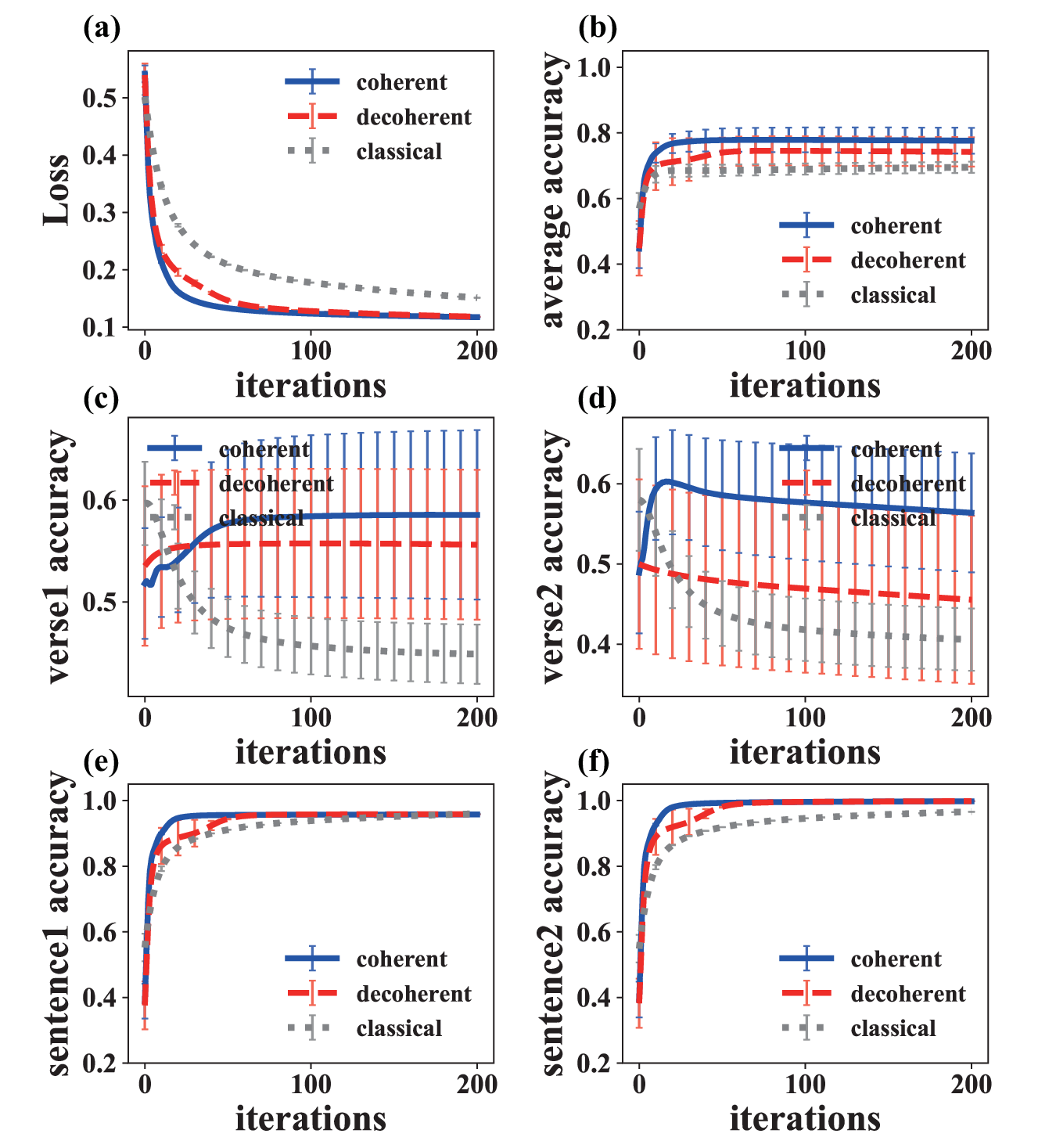}\\
          \caption{As introduced in Sec. \ref{subsec:accelerate}, the loss function of the coherent QSNN converges the fastest (see a).
          The coherent QSNN performs best on the test set (see b).
          And the main advantage of the coherent QSNN is that it can recognize verse1 (see c) and verse2 (see d) with relatively high accuracy.
          Meanwhile, in recognizing 2 normal sentences (see e and f), the QSNNs win only a little since the recognition accuracy of all three kinds of networks is already close to 1. The error bars are drawn as the sample variances.}
          \label{fig:verse-recognition}
        \end{figure}

        In this task, the QSNNs consist of 8 neurons $\{|1\rangle,|2\rangle,\cdots,|8\rangle\}$ in the hidden layer, which correspond to 8 words.
        The training set we use here consists of 12 labeled word sequences made up of those 8 words and some stop words.
        The test set consists of 2 verses and 2 normal sentences, whose probabilities of being recognized as a sentence (Yes) (i.e. the accuracy) are used to evaluate the performance of the trained QSNNs (both coherent and decoherent) and classical NN.
        We train the coherent QSNN, decoherent QSNN, and the classical NN from 15 random initialization samples and average over the samples.
        More details about the training and test are given in Appendix \ref{appendix:details about verse}.

        The training and test results are shown in Fig.\,\ref{fig:verse-recognition}.
        As shown in Fig.\,\ref{fig:verse-recognition}(a), the loss function of the coherent QSNN goes down the fastest, which is consistent with the result of Sec. \ref{subsec:accelerate}, namely, the QSNN has the advantage of reducing the number of training steps.
        Fig.\,\ref{fig:verse-recognition}(b) shows that the coherent QSNN gives the highest average accuracy over all word sequences in the test set.
        As shown in Fig.\,\ref{fig:verse-recognition}(c-f), the recognition accuracy of all three networks for verses is lower than that for normal sentences because there are no verses in the training set, so the word order of the verses is unusual for the networks.
        However, two QSNNs have an improvement in the accuracy of verse recognition compared with classical NN (see Fig.\,\ref{fig:verse-recognition}(c-d)).
        And the coherent QSNN achieves the highest accuracy of verse recognition.
        Meanwhile, in recognizing 2 normal sentences (see Fig.\,\ref{fig:verse-recognition}(e-f)), the QSNNs win only a little, because the recognition accuracy of all three kinds of networks is already close to 1.
        In summary, when we extend the networks to 11 neurons, the result in Sec. \ref{subsec:accelerate} that the QSNN requires fewer training steps still holds.
        Moreover, the QSNN, especially the coherent one, improves the accuracy of verse recognition without damaging its accuracy in recognizing normal sentences.

        In order to exclude the possibility that the particularity of the training set affects the results, we use another data set to complete the same task. The result is shown in Appendix \ref{appendix:details about verse}.

    \section{Robustness}\label{sec:robustness}
        \subsection{Label noise}\label{subsec:label noise}
        Our QSNN aiming at accurate classification is based on supervised learning, which has the implicit assumption that annotators are experts and they would provide perfectly labeled training data.
        It is however rarely true in real-world scenarios. Some of the labels in the provided training set may be wrong.
        And the development of grammar also may lead to changes in the labels of some word sequences in the matter of our concern, which increases the mislabeled data (i.e. label noise) inevitably.
        When given a QSNN trained with a noisy training set, it is considered robustness if its loss function can be optimized to the optimum with a correct training set faster than the alternatives.

        In this subsection, we investigate the robustness of two QSNNs (i.e. coherent QSNN and decoherent QSNN) against label noise in sentence recognition.
        The structure of the QSNN and correct training set used here are the same as that in Sec. \ref{sub:verse-recognition}.
        Because the coherent QSNN performs better than the decoherent one if they are initialized randomly, as shown in Fig.\,\ref{fig:verse-recognition}(a), we train both of them from 4 uniform initialization samples $\{h_i=0.1, \gamma_j=0.1, 0.3, 0.5, 0.7\}_{i,j=1, 1}^{m, n}$ to exclude the effect of the random initialization, where $m$ ($n$) is the number of parameters in the Hamiltonian (Lindblad operators).

\begin{figure}[h]
          \centering
          \includegraphics[width=8.5cm]{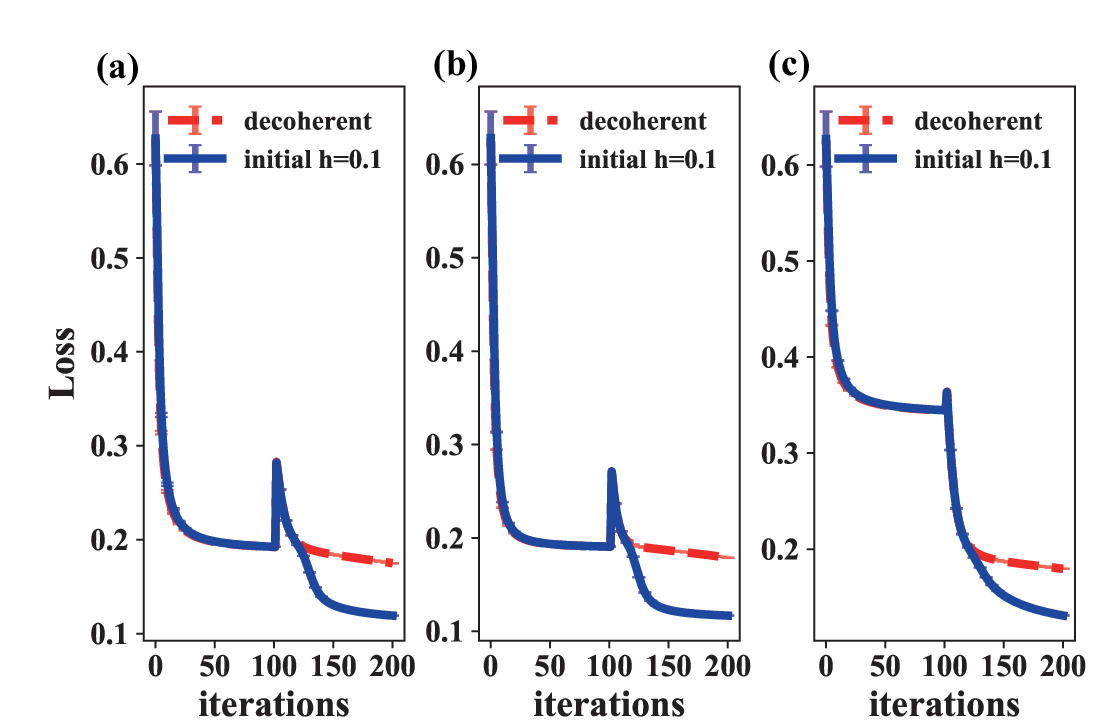}\\
          \caption{We show the loss functions (the mean of 4 uniform initialization samples) of the coherent and decoherent QSNN with respect to the number of iterations used in the training procedure.
          The sample mean Loss of both networks is indistinguishable at first, but the one given by the coherent QSNN shows a better convergence after the errors are corrected at the 100th iteration ((a), (b), and (c) correspond to three different corrupted sets).
          More training details are given in Appendix \ref{appendix:details about noise}. The error bars are drawn as the sample variance.}
          \label{fig:label-noise}
        \end{figure}

        In the first 100 iterations, we train the two kinds of QSNNs with three corrupted training sets consisting of different wrong labels.
        The results corresponding to the three corrupted sets are shown in (a), (b), and (c) of Fig.~\ref{fig:label-noise}.
        Then, the wrong labels are corrected in the 100th iteration, and the QSNNs are trained with the correct training set in the later iterations.
        As shown in Fig.~\ref{fig:label-noise}, the loss functions (the mean of 4 initialization samples) of two kinds of QSNNs are indistinguishable before the 100th iteration, but the coherent QSNN's loss function converges faster after the errors are corrected.
        In terms of these results, the coherent QSNN with 11 neurons is more robust against the label noise than the decoherent one.

    \subsection{Device noise}\label{subsec:device noise}

    In our approach, we update the parameters of the QSNN in training to find the optimal values of the parameters.
    However, the values of parameters may not be set precisely in practical application because of the presence of device noise, and minor errors in parameter values may affect the performance of the QSNN.
    So, an ideal network needs to be robust to perturbations of the parameters caused by device noise.
    In this subsection, we compare the robustness of the coherent and decoherent QSNN against the device noise in sentence recognition.

\begin{figure}[h]
      \centering
      \includegraphics[width=8.5cm]{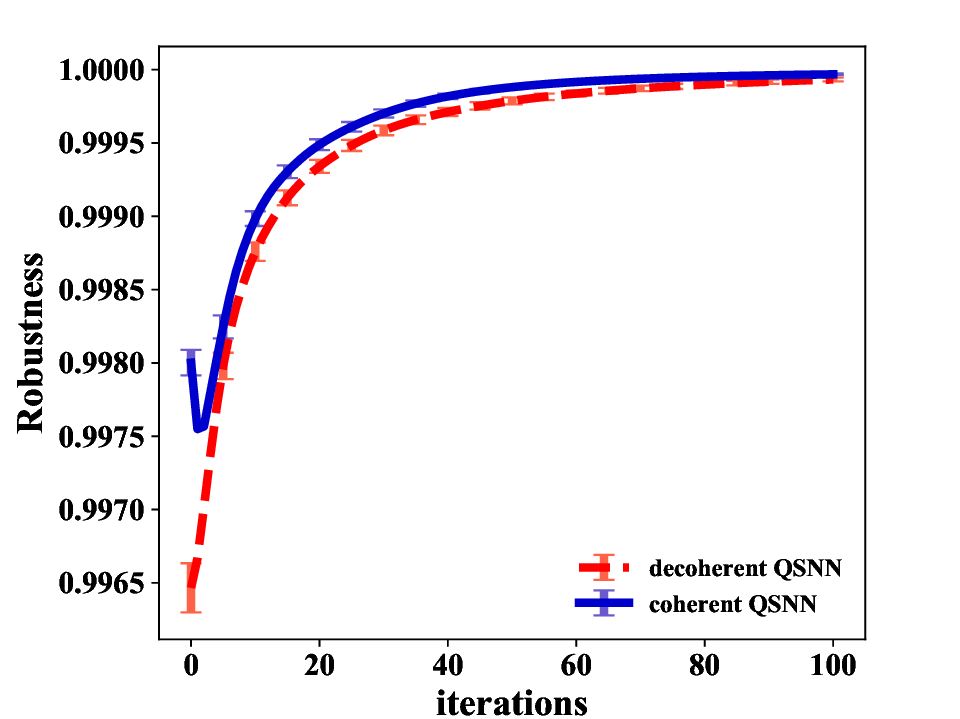}\\
      \caption{We compute the Robustness in the training process described in Sec. \ref{subsec:accelerate}. The robustness of the coherent QSNN is higher than the decoherent one.
      The error bars are drawn as 95\% confidence intervals.}
      \label{fig:robsutness}
    \end{figure}

    There is the same number of Lindblad parameters $\gamma_k$ in both the coherent and decoherent QSNN.
    So, we consider the effect of imprecise setting of these Lindblad parameters on robustness here.
    The QSNN gives the output $\rho_{\text{out}}$ for the $s$th input sample. One can perform a corresponding measurement $\Omega^s$ on the output to obtain the successful classification probability $\text{Tr}(\Omega^s\rho_{\text{out}}^s)$ of the $s$sample.
    We use the derivative $\frac{\partial \left[\text{Tr}(\Omega^s\rho_{\text{out}}^s)\right]}{\partial\gamma_k}$ to describe the effect of the perturbation $\delta\gamma_k$ on the successful classification probability of the $s$sample. And the robustness for a batch of $M$ samples is defined as
    \begin{equation}\label{eq:robustness}
    \mbox{Robustness}=1-\frac{1}{M}\frac{1}{n}\sum_{s=1}^M\sum_{k=1}^n|\frac{\partial \left[\text{Tr}(\Omega^s\rho^s)\right]}{\partial\gamma_k}|^2,
    \end{equation}
    where $n$ is the number of Lindblad parameters $\gamma_k$.
    We take the training process described in Sec. \ref{subsec:accelerate} as an example to calculate the robustness of two kinds of QSNNs in the training process.
    As shown in Fig.\,\ref{fig:robsutness}, the robustness of the coherent QSNN is higher than the decoherent one.
    Namely, the coherent QSNN is more robust against device noise.
    So, in the sentence recognition task, the advantages of the coherent QSNN over the decoherent one displayed in Sec. \ref{sec:quantum advantages} are not at the cost of robustness.

    \section{Conclusion and outlook}

    We study a quantum stochastic neural network (QSNN) based on quantum stochastic walks and use a learning algorithm based on gradient descent to train it.
    Given an input, the trained QSNN can correspondingly give a prediction that is approximate to the target output.
    Our numerical simulation shows that the toy models of the QSNN have abilities to accomplish classical tasks such as function approximation, 2D data classification, and sequence classification.
    We investigate the performance of QSNN, taking sentence recognition, a specific sequence classification task, as an example.
    Considering a toy model with 5 neurons, we find that our QSNNs (especially the one with coherent evolution) reduce the number of training steps.
    And when 11 neurons are implemented, we find that our QSNNs (especially the one with coherent evolution) not only have an advantage in training speedup, but also in improving the accuracy of recognizing new types of inputs like the verses.
    Meanwhile, the coherent QSNN in our simulations of the sequence recognition task also performs better than the decoherent QSNN in terms of robustness against label noise and device noise.
    All these results suggest that QSNNs, especially the coherent QSNN, have certain quantum advantages over classical neural networks. These quantum advantages might be intrinsically related to the advantages shown in shadow tomography.
    Whether these advantages still exist in more complex quantum neural networks is worth further exploration.

    We should also be aware that we use classical numerical simulation to calculate the gradients in order to update the network parameters, while in fully quantum implementation of our neural network the quantum operations to evaluate the gradients are highly nontrivial and need further exploration.
    An efficient universal learning algorithm purely based on quantum operations, as the quantum counterpart of the backpropagation algorithm, is a prerequisite to fully unlock the quantum advantages of quantum neural networks.

    Some recent theoretical proposals about quantum neural networks based on quantum gates  \cite{wan2017quantum,beer2020training, farhi2018classification, mitarai2018quantum,bondarenko2020quantum} are experimentally demanding so far.
    However, some experimental advances of quantum walks \cite{du2003experimental, tang2018experimental, perets2008realization, tang2019experimental, laneve2021experimental} make our QSNN promising for implementation.

    In this work, our toy model of QSNN has shown certain advantages in an elementary example of sentence recognition.
    We hope it can be developed and applied to more sophisticated tasks in the future.

	\section*{Acknowledgments}
This work is supported by the National Key R\&D Program of China (Grant No. 2017YFA0303703)
and the National Natural Science Foundation of China (Grant No. 12175104).

\appendix
    \section{Learning algorithm}\label{appendix:learning algorithm}

        In the main text, the QSNNs are applied to 3 tasks with different structures and loss functions.
        In this part, we discuss how to use gradient descent to optimize the parameters of the QSNNs in the different tasks.

        First, in the task of function approximation, there are $M$ samples consisting input $x_s$ and target output $f(x_s)$ in the training set $\{(x_s, f(x_s))\in \mathbb{R}^2)\}_{s=1}^M$.
        The $s$th input data $x_s$ are encoded in the initial state of the QSNN $\rho_\text{in}^s$ according to Eq.\,(\ref{eq:encode}).
        The QSNN only contains the parameter $\vec{h}$ characterizing the coherent evolution $\Lambda(\vec{h})$.
        The state of the QSNN after above evolution $\Lambda(\vec{h})$ for the $s$th sample can be given as
        \begin{equation}
        |\rho_{\text{out}}^s\rangle=\Lambda(\vec{h})|\rho_{\text{in}}^s\rangle=e^{\mathcal{L}(\vec{h})T}|\rho_{\text{in}}^s\rangle
        \end{equation}
        by solving Eq. (\ref{eq:cj master equation}), where $T$ is the evolution time.
        One can perform a projection onto the output neuron by measuring $\Omega=|N-1\rangle\langle N-1|$ on the final state $\rho_{\text{out}}^s$ to give the output $\text{Tr}(\Omega\rho_{\text{out}}^s)$.
        The loss function for a batch of $M$ samples is defined as
        \begin{equation}
        \mbox{Loss}=\frac{1}{M}\sum_{s=1}^M\left[\text{Tr}(\Omega\rho_{\text{out}}^s)-f(x_s)\right]^2.
        \end{equation}
        The gradient of the loss function with respect to any parameter $\theta$ can be written as
        \begin{equation}\label{eq:gradient_1}
        \frac{\partial \mbox{Loss}}{\partial \theta}=\frac{1}{M}\sum_{s=1}^M2\left[\text{Tr}(\Omega\rho_{\text{out}}^s)-f(x_s)\right]\text{Tr}(\Omega\frac{\partial\rho_{\text{out}}^s}{\partial \theta}).
        \end{equation}
        Then, the gradient of the final state with respect to any parameter $\theta$, which can be given as
        \begin{equation}
        \frac{\partial|\rho_{\text{out}}^s\rangle}{\partial h_k}=\frac{\partial \Lambda(\vec{h})}{\partial h_k}|\rho_{\text{in}}^s\rangle,
        \end{equation}
         where
        \begin{equation}
        \frac{\partial \Lambda(\vec{h})}{\partial h_k}=\int_0^T dt\left[e^{\mathcal{L}(T-t)}\frac{\partial \mathcal{L}(\vec{h})}{\partial h_k}e^{\mathcal{L}t}\right].
        \end{equation}
        The partial deviations of the operator $\mathcal{L}$ with respect to parameters can be obtained through simple derivation
         \begin{equation}
         \frac{\partial\mathcal{L}}{\partial h_k}=-i(\mu_k\otimes I-I\otimes \mu_k^T)
         \end{equation}
        where $\mu_k=\frac{\partial H}{\partial h_k}$.

        Second, we introduce the gradient calculation in the classification tasks (classification of 2D data and sequence classification).
        The classical data can be encoded in the state of the networks $\rho_{\text{in}}^s$ for the $s$th sample.
        Then, the evolution of the QSNNs in two tasks can both involve two stages: \emph{Unitary evolution} $\Lambda^U(\vec{h})$ and \emph{Dissipative evolution} $\Lambda^D(\vec{\gamma})$, and the final state can be given as
        \begin{equation}
        \begin{aligned}
        |\rho_{\text{out}}^s\rangle&=\Lambda^D(\vec{\gamma})|\rho_U\rangle=\Lambda^D(\vec{\gamma})\Lambda^U(\vec{h})|\rho_{\text{in}}^s\rangle\\
        &=e^{\mathcal{L}_D(\vec{\gamma}) T^D}e^{\mathcal{L}_H(\vec{h}) T^U}|\rho_{\text{in}}^s\rangle
        \end{aligned}
        \end{equation}
        by solving Eq. (\ref{eq:cj master equation}), where $T^U$ and $T^D$ are the evolution time of the \emph{Unitary evolution} and \emph{Dissipative evolution}, respectively.
        One can perform the projection $\Omega^s=|l_s\rangle\langle l_s|$ corresponding to the label of the $s$th sample onto the output layer to give the probabilities of correct classification $\text{Tr}(\Omega^s\rho_{\text{out}}^s)$.
        The loss function is defined as
        \begin{equation}
        \text{Loss}=1-\frac{1}{M}\sum_{s=1}^M\text{Tr}(\Omega^s\rho_{\text{out}}^s).
        \end{equation}
        The gradient of the loss function with respect to any parameter $\theta$ can be given as
        \begin{equation}\label{eq:gradient_2}
        \frac{\partial \text{Loss}}{\partial \theta}=-\frac{1}{M}\sum_{s=1}^M\text{Tr}(\Omega^s\frac{\partial\rho_{\text{out}}^s}{\partial \theta})
        \end{equation}
        Then, the gradients of the final state with respect to the parameters of the Hamiltonian ($h_k$) and Lindblad operators ($\gamma_k$) can be separately given as
        \begin{equation}
        \begin{aligned}
        \frac{\partial|\rho_{\text{out}}^s\rangle}{\partial h_k}&=\Lambda^D(\vec{\gamma})\frac{\partial \Lambda^U(\vec{h})}{\partial h_k}|\rho_{\text{in}}^s\rangle\\
        \frac{\partial|\rho_{\text{out}}^s\rangle}{\partial \gamma_k}&=\frac{\partial\Lambda^D(\vec{\gamma})}
        {\partial\gamma_k}\Lambda^U(\vec{h})|\rho_{\text{in}}^s\rangle,
        \end{aligned}
        \end{equation}
        where
        \begin{equation}
        \begin{aligned}
        \frac{\partial \Lambda^U(\vec{h})}{\partial h_k}&=\int_0^{T^U}dt\left[e^{\mathcal{L}_U(T^U-t)}\frac{\partial \mathcal{L}_H(\vec{h})}{\partial h_k}e^{\mathcal{L}_Ut}\right]\\
        \frac{\partial \Lambda^D(\vec{\gamma})}{\partial \gamma_k}&=\int_0^{T^D}dt\left[e^{\mathcal{L}_D(T^D-t)}\frac{\partial \mathcal{L}_D(\vec{\gamma})}{\partial \gamma_k}e^{\mathcal{L}_Dt}\right].
        \end{aligned}
        \end{equation}
         The partial deviations of the operator $\mathcal{L}$ with respect to parameters can be obtained through simple derivation
         \begin{equation}
         \begin{aligned}
         \frac{\partial\mathcal{L}_H}{\partial h_k}&=-i(\mu_k\otimes I-I\otimes \mu_k^T)\\
         \frac{\partial\mathcal{L}_D}{\partial\gamma_k}&=2\gamma_k\left[\nu_k\otimes \nu_k^*-\frac{1}{2}(\nu_k^\dagger\nu_k )\otimes I-\frac{1}{2}I \otimes(\nu_k^T\nu_k^*)\right],
         \end{aligned}
         \end{equation}
        where $\mu_k=\frac{\partial H}{\partial h_k}$ and $\nu_k=\frac{\partial L_k}{\partial \gamma_k}$.

        After the calculation of the gradient described in Eqs.\,(\ref{eq:gradient_1}) and (\ref{eq:gradient_2}), the parameter can be updated according to the update rule
        \begin{equation}
        \theta^\prime=\theta-\eta\frac{\partial \mbox{Loss}}{\partial\theta}
        \end{equation}
        to minimize the \mbox{Loss}, where $\eta$ is an adjustable non-negative parameter called the learning rate.

	\section{QSNN's expression ability in sentence recognition}\label{appendix:expression}

\begin{figure}[ht]
          \centering
          \includegraphics[width=8cm]{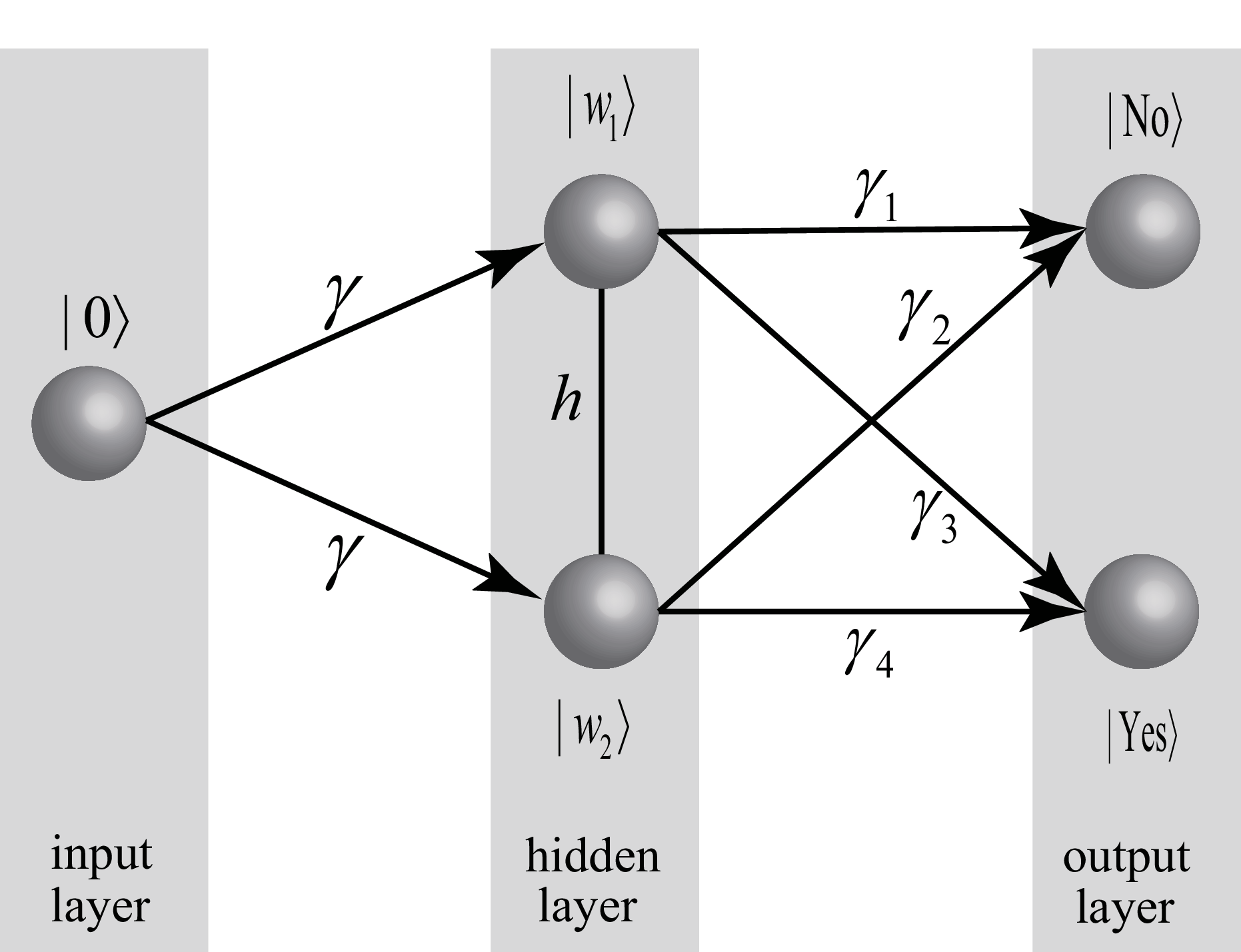}
          \caption{The QSNN structure for the case with only 2 words in our corpora. The QSNN is initialized in the state $|0\rangle$ (representing the single neuron in the input layer).
          The two neurons represented by $|w_1\rangle=|1\rangle$ and $|w_2\rangle=|2\rangle$ in the hidden layer correspond to the two different words $w_1, w_2$ in the corpora.
          The output layer tells whether the input sequence is a sentence $(|\text{Yes}\rangle)$ or not $(|\text{No}\rangle)$. $h\in\mathbb{R}$ characterizes the coherent strength between two neurons in the hidden layer, and $\{\gamma_k\in\mathbb{R}\}_{k=1}^4$ characterize the dissipation rates from the neurons in the hidden layer to those in the output layer.}
          \label{fig:5dim}
        \end{figure}

        We have shown how to train the QSNN to classify sequences with a training set containing 2 different objects $\{w_1, w_2\}$ in the main text.
        In this part, we take the sentence recognition task as an example to investigate the expression ability of a QSNN (as shown in Fig.\,\ref{fig:model}(c)) in the sequence classification task.
        Sentence recognition here is to classify word sequences according to whether they are grammatical sentences (Yes) or not (No).
        Fig.~\ref{fig:5dim} shows more details about the QSNN we studied here.
        The training set is donated as $S=\{(e_s, l_s)\}_{s=1}^2$, where the $s$th input (a sequence of the two words) $e_s=(w_i, w_j)_{i\neq j\in\{1,2\}}$ is labeled $l_s\in\{\text{Yes}, \text{No}\}$ according to whether it is a grammatical sentence or not.
        For a certain input sample, the QSNN can involve an \emph{Input} stage, a \emph{Unitary evolution}, a \emph{Dissipative evolution}, and a final \emph{measurement} stage, as described in the main text.

        Now we demonstrate that a simple QSNN (as shown in Fig.\,\ref{fig:model}(c)), even without the \emph{Unitary evolution}, can be trained to correctly classify any input sequences with high probabilities here.

        After the \emph{Input} process governed by the dissipation rate $\gamma$ for the sequence $(w_1, w_2)$, the state of the network is given as Eq. (\ref{eq:rho_in}), namely
        \begin{equation*}
        \rho_{\text{in}}=\omega|0\rangle\langle0|+\alpha|1\rangle\langle1|+\beta|2\rangle\langle2|,
        \end{equation*}
        where $\omega=e^{-3\gamma^2}$, $\beta=1-e^{-\gamma^2}+(e^{-\gamma^2}-e^{-3\gamma^2}) /2$, $\alpha=(e^{-\gamma^2}-e^{-3\gamma^2})/2 $.
        The evolution time $T^{\text{in}}$ of the \emph{Input} is set as 1.
        If the input sequence changes from $(w_1, w_2)$ to $(w_2, w_1)$, the value of $\omega$ doesn't change, but the values of $\alpha$ and $\beta$ are swapped.
        Then, there is an \emph{Dissipative evolution} process described by a decoherent evolution $\Lambda^D(\vec{\gamma})$ for a duration $T^D$, where $\vec{\gamma}=(\gamma_1, \gamma_2, \gamma_3, \gamma_4)\in\mathbb{R}^4$ is the dissipation rate.
        After the \emph{Dissipative evolution} process, the final state of the QSNN can be written as $\rho_{\text{out}}$.
        Performing a projective measurement $|N-1\rangle\langle N-1|$ on $\rho_{\text{out}}$ gives Yes (the input is a grammatical sentence) with a probability
        \begin{equation*}
        \begin{aligned}
        p_{\text{Yes}}&=\langle \text{Yes} | \rho_{\text{out}} | \text{Yes} \rangle\\
        &=\frac{1-e^{-(\gamma_2^2+\gamma_4^2)}}{\gamma_2^2+\gamma_4^2}\gamma_4^2\beta+
        \frac{1-e^{-(\gamma_1^2+\gamma_3^2)}}{\gamma_1^2+\gamma_3^2}\gamma_3^2\alpha.
       \end{aligned}
       \end{equation*}
       And performing a projective measurement $|N-2\rangle\langle N-2|$ on $\rho_{\text{out}}$ gives No (the input is not a grammatical sentence) with a probability
        \begin{equation*}
        \begin{aligned}
        p_{\text{No}}&=\langle \text{No} | \rho_{\text{out}} | \text{No} \rangle\\
        &=\frac{1-e^{-(\gamma_2^2+\gamma_4^2)}}{\gamma_2^2+\gamma_4^2}\gamma_2^2\beta+
        \frac{1-e^{-(\gamma_1^2+\gamma_3^2)}}{\gamma_1^2+\gamma_3^2}\gamma_1^2\alpha,
        \end{aligned}
        \end{equation*}
        noting that we have set the dimensionless $T^D=1$ in the calculation.
        Just for clarity, we consider the difference
        $$p_{\text{No}}-p_{\text{Yes}}=g(\gamma_2,\gamma_4)\beta+ g(\gamma_1,\gamma_3)\alpha,$$ which is close to $-1$ if the sequence is a sentence (Yes) and 1 if it isn't (No).
        If the values of $\gamma$ and $T^D$ are sufficiently large, we can always find that the probability $1-p_{\text{Yes}} -p_{\text{No}}$ is negligible and the probability distribution $(\alpha, \beta)$ over neurons $|w_1\rangle$ and $|w_2\rangle$ (see Fig.\,\ref{fig:5dim}) close to $(1, 0)$ for input $e_s=(w_1, w_2)$ or close to $(0, 1)$ for input $e_s=(w_2, w_1)$.
        \begin{figure}[ht]
          \centering
          \includegraphics[width=9.5cm]{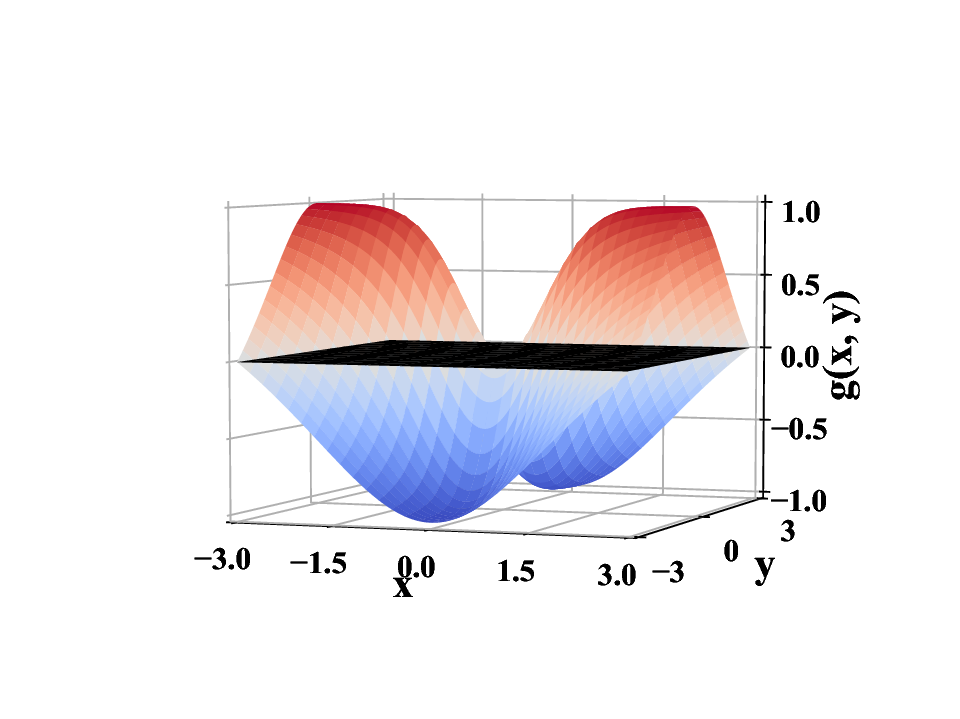}\\
          \caption{The relationship between the function $g$ and the argument $x, y$ (representing $\gamma_2, \gamma_4$ or $\gamma_1, \gamma_3$).}\label{fig:expression}
        \end{figure}

         The surface in Fig.\,\ref{fig:expression} represents the relationship between the function $g$ and the argument $x, y$ (representing $\gamma_2, \gamma_4$ or $\gamma_1, \gamma_3$).
         If the input is $(w_1, w_2)$, then $p_{\text{No}}-p_{\text{Yes}}$ is close to $g(\gamma_1,\gamma_3)$, and a point $(\gamma_1,\gamma_3)$ can be found on the surface to make $p_{\text{No}}-p_{\text{Yes}}\rightarrow1$ (the input is not a sentence) or $p_{\text{No}}-p_{\text{Yes}}\rightarrow-1$ (the input is a sentence).
         If the input is $(w_2, w_1)$, then $p_{\text{No}}-p_{\text{Yes}}$ is close to $g(\gamma_2,\gamma_4)$, and a point $(\gamma_2,\gamma_4)$ also can be found on the surface for $p_{\text{No}}-p_{\text{Yes}}\rightarrow1$ (the input is not a sentence) or $p_{\text{No}}-p_{\text{Yes}}\rightarrow-1$ (the input is a sentence).

        In summary, the optimal solution for the values of the dissipation rates $\vec{\gamma}=(\gamma_1,\gamma_2, \gamma_3, \gamma_4)\in\mathbb{R}^4$ exists to classify any possible training sample with a high success probability.

    \section{Training and test details}
        In this appendix, we provide some details about the simulation described in Secs. \ref{sec:quantum advantages} and \ref{sec:robustness}.
        In those simulations, the dimensionless evolution time is set as $T^{I}=T^{U}=T^{D}=20$, the Lindblad parameters (will not be updated) governing the \emph{Input} process is set as $\gamma=1$, and the learning rate decreases with the number of iterations.
        The learning rate corresponding to the parameter $\theta$ can be written as $\eta^{\theta}=\eta_0^{\theta}/(1+\frac{n_{\text{iteration}}}{R})$, where $\eta_0^{\theta}$ is the initial learning rate set at the beginning of the training, $n_{\text{iteration}}$ is the number of iterations, and $R=15$ controls the decay rate $\frac{1}{R}$ of the learning rate.
        Then, the specific details corresponding to each simulation are shown in the corresponding subsections below.

    \subsection{training performance}\label{appendix:details about accelerate}

         We numerically simulate the training of three types of neural networks, namely, the coherent QSNN, the decoherent QSNN, and the classical analogous named classical NN with gradient descent in sentence recognition.
         The result is shown in Fig.\,\ref{fig:accelerate fitting}. Here, we provide some simulation details.

        We train the QSNNs from 1000 random initialization samples $\vec{\gamma}\in\mathbb{R}\ (\gamma_i\in[-1, 1]\ \text{uniformly})$ and average over the samples.
        Here, the initial learning rates are set as $\eta_0^h=0.1$ and $\eta_0^{\vec{\gamma}}=\eta_0^{\vec{w}}=\eta_0^{\vec{b}}=1$.
        The result is shown in Fig.\,\ref{fig:accelerate fitting}, where we plot the sample mean Loss of the three networks (i.e., the classical NN (grey), decoherent QSNN (red), and coherent QSNN with initial $h=0.1$ (blue)) with respect to the number of iterations used in the training procedure.
        The classical NN is initialized by random $\vec{w}, \vec{b}\ (w_i, b_i\in[-1,1]\ \text{uniformly})$.

        \begin{figure}[ht]
          \centering
          \includegraphics[width=8.5cm]{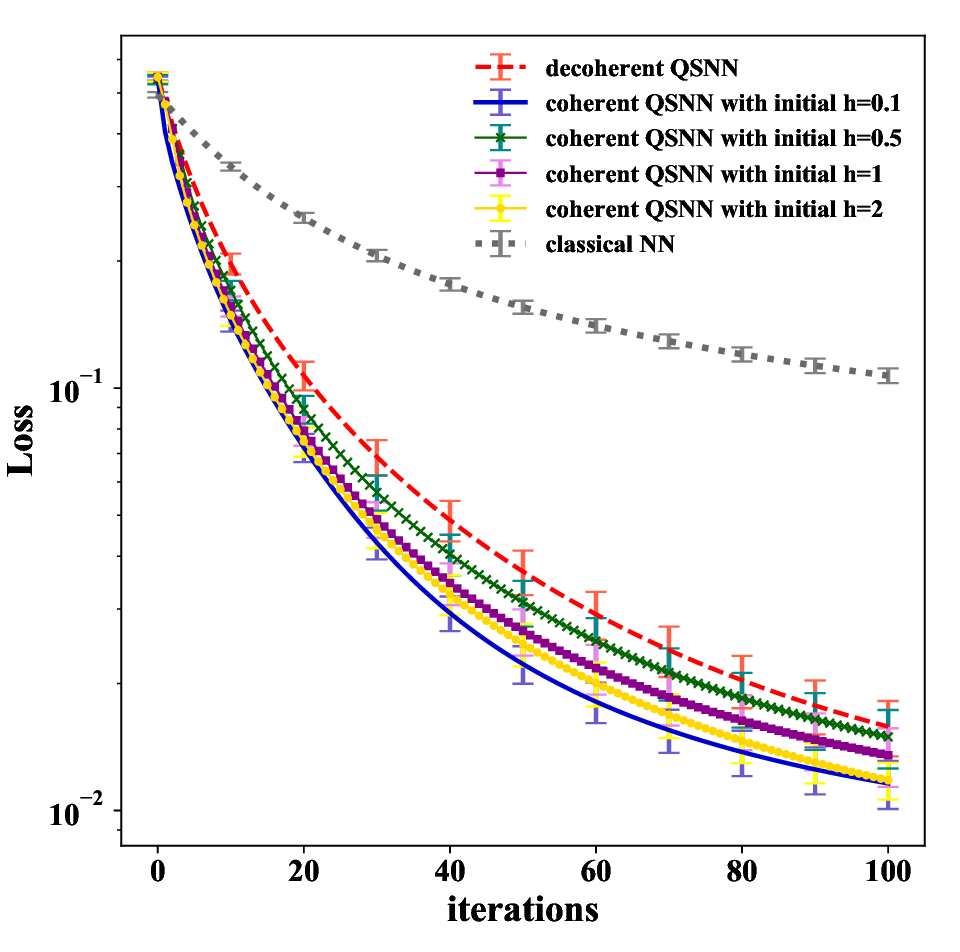}\\
          \caption{A plot of the loss function (the mean of 1000 initialization samples) of the networks with respect to the number of iterations used in the training procedure.
          Different curves in the figure respectively represent the results of different networks (see legend).
          The loss functions of the coherent QSNNs always converge faster than the decoherent QSNN and classical NN.
          But the loss functions of the coherent QSNNs with different initial values of $h$ converge at different speeds.
          The error bars are drawn as 95\% confidence intervals.}
          \label{fig:different h}
        \end{figure}

        We find the initial value of the coherent strength $h$ has a certain influence on the training performance of the coherent QSNN.
        So, we train the coherent QSNNs with different initial values $\{0.1, 0.5, 1, 2\}$ of $h$ from 1000 random initialization samples $\vec{\gamma}\in\mathbb{R}\ (\gamma_i\in[-1, 1]\ \text{uniformly})$.
        We give the training results in Fig.\,\ref{fig:different h}.
        As shown in Fig.\,\ref{fig:different h}, the training steps required by the coherent QSNN can be fewer when we choose an appropriate initialization than a random one.
        Then, we provide a strategy to choose an appropriate initial value for the coherence strength $h$ in the light of experience.
        One can compute the values of Loss of the coherent QSNNs with different initial values of $h$ before training.
        The most appropriate initial value of $h$ can be directly chosen as the value that gives the minimum initial sample Loss in the previous computation.
        \begin{figure}[ht]
          \centering
          \includegraphics[width=8.5cm]{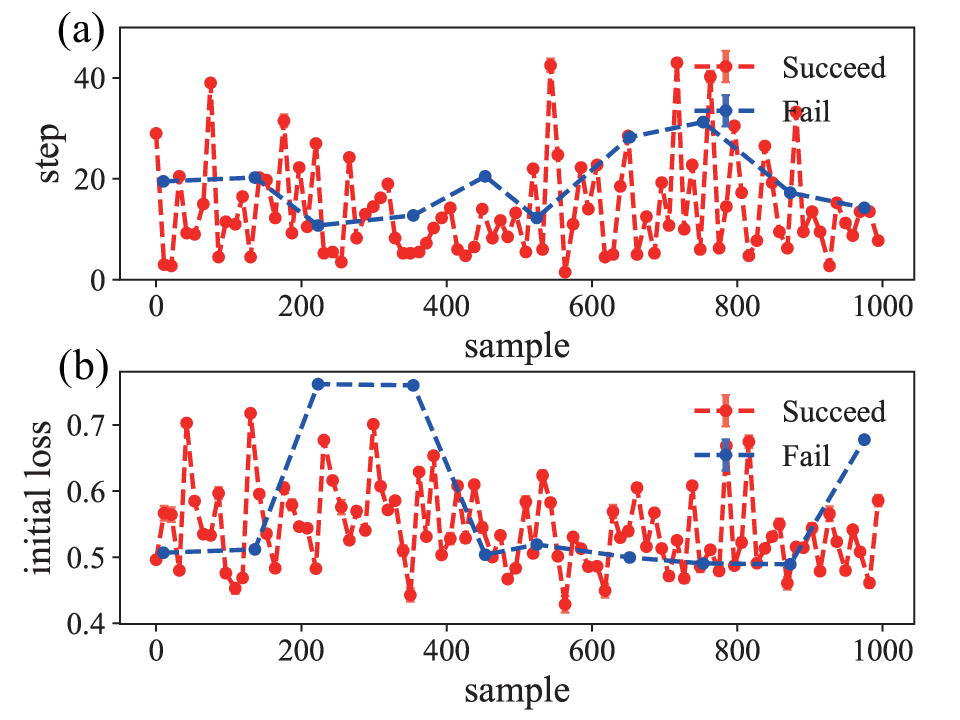}\\
          \caption{We compute the mean convergence step (a) and the mean initial loss (b) for the 1000 initialization samples $\vec{\gamma}\in\mathbb{R}\ (\gamma_i\in[-1, 1]\ \text{uniformly})$ whose initial coherent strengths $h$ can be chosen from $\{0.1, 0.5, 1, 2\}$.
          The convergence step is determined by the minimum number of iterations required for the QSNN to make $\text{Loss}<0.1$.
          The initial loss is the value of Loss given by the QSNN before training.
          We denote the set of $h$ values which give the minimum initial Loss as $S_{\min(\text{Loss})}$, and denote the set of $h$ values which give the least convergence steps, as $S_{\min(\text{step})}$.
          If $S_{\min(\text{Loss})}\cap S_{\min(\text{step})}=S_{\text{success}}\neq \varnothing$, it is believed that the strategy works and we draw the point in red. If $S_{\min(\text{Loss})}\cap S_{\min(\text{step})}=\varnothing$, the strategy doesn't work and we draw the point in blue. The error bars are drawn by the standard deviations of different $h$ values samples.}
          \label{fig:find-best-h}
        \end{figure}

        We use the convergence step to quantify the convergence speed of the coherent QSNN.
        It is determined by the minimum number of iterations required for the QSNN to make $\text{Loss}<0.1$.
        For each initialization samples $\vec{\gamma}\in\mathbb{R}$, we compute the convergence steps and the initial loss (i.e., values of Loss of a network with initial parameters) of 4 QSNNs with different $h$, and average over the 4 QSNNs.
        The average value of convergence steps and initial loss for each sample are shown in Fig.\,\ref{fig:find-best-h}(a) and (b), respectively.
        We denote the set of $h$ initial values which make the QSNNs give the minimum initial Loss as $S_{\min(\text{Loss})}$, and denote the set of $h$ initial values which make the QSNNs give the least convergence steps, as $S_{\min(\text{step})}$.
        If $S_{\min(\text{Loss})}\cap S_{\min(\text{step})}=S_{\text{success}}\neq \varnothing$, it is believed that the minimum convergence step (the maximum convergence speed) can be given by the $h$ which also gives minimum initial Loss.
        Namely, the strategy we proposed above works, then we draw the point (in Fig.\,\ref{fig:find-best-h}) of the sample in red. If $S_{\min(\text{Loss})}\cap S_{\min(\text{step})}=\varnothing$, the strategy doesn't work and we draw the point in blue.
        We can choose the appropriate initial $h$ from $S_{\text{success}}$ when the strategy works.
        Fig.\,\ref{fig:find-best-h} shows that the strategy works ($S_{\text{success}}\neq\varnothing$) with 90\% probability of success.

    \subsection{verse recognition}\label{appendix:details about verse}
        In Sec. \ref{sub:verse-recognition}, we compare the test performances of the coherent QSNN, decoherent QSNN, and classical NN, and find the first one performs best in the verse recognition. We just provide some training and test details here. In this case, there are 8 neurons in the hidden layer corresponding to the 8 different words in the corpus.
        The training and test set are listed in TABLE \ref{tab:set}.
        In our simulation, before word sequences are input into the neural networks, all input words in them are lemmatized to their normal forms, and stop words are removed from the word sequences.
        We compute the mean Loss and mean accuracy of the networks trained from 15 random initialization QSNN samples $\vec{\gamma}$ or classical NN samples $\vec{w},\vec{b}\ (\gamma_i,w_i,b_i\in[-1, 1]\ \text{uniformly})$.
        Here, the initial learning rates are chosen as $\eta_0^{\vec{h}}=0.5$ and $\eta_0^{\vec{\gamma}}=\eta_0^{\vec{w}}=\eta_0^{\vec{b}}=3$.
        When the number of iterations $n_{\text{iteration}}>100$, all the learning rates no longer decrease in the simulation of this section.

\begin{figure}[h]
          \centering
          \includegraphics[width=8.5cm]{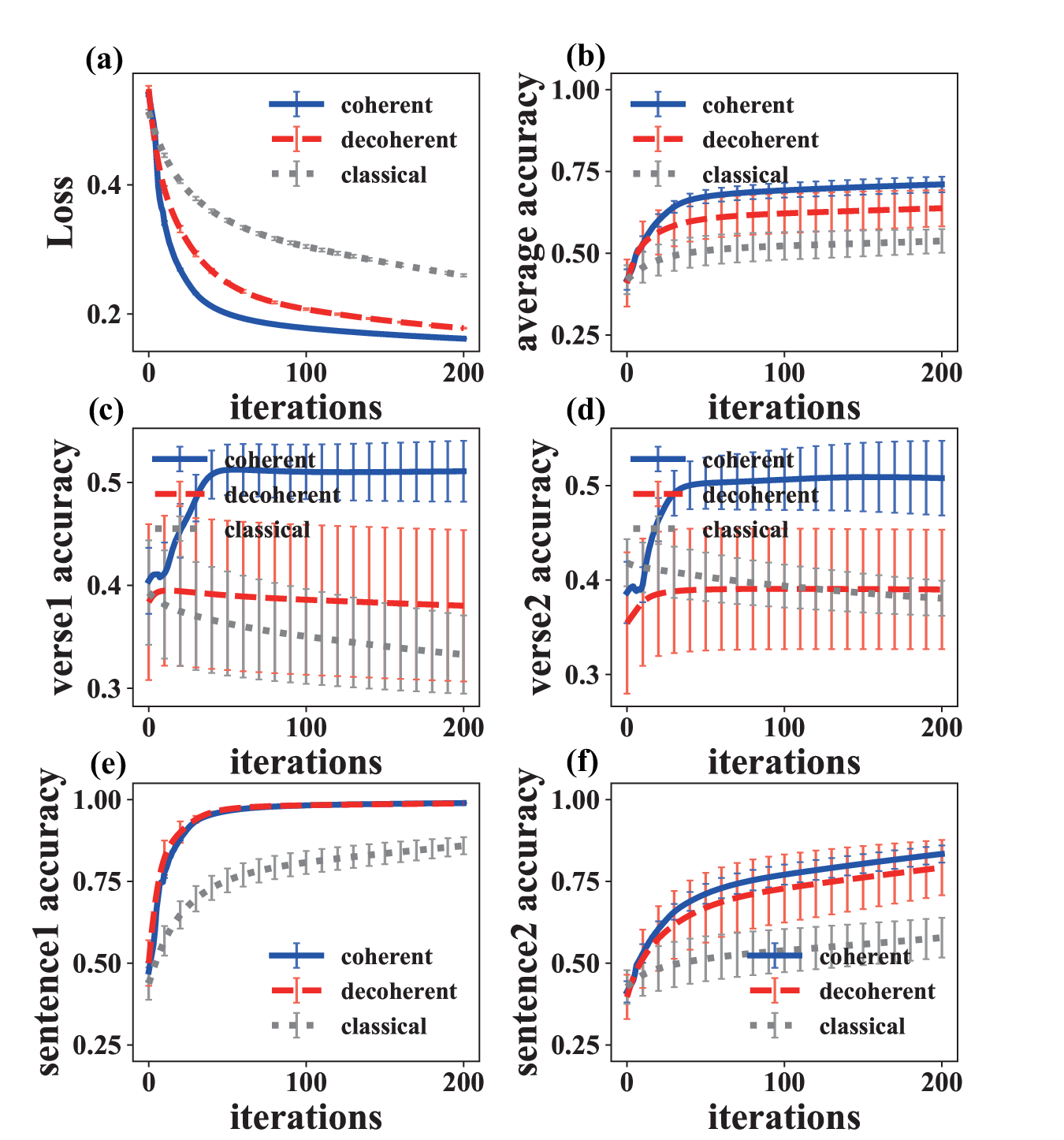}\\
          \caption{Training and test performance of the new corpora.
          The loss function of the coherent QSNN converges the fastest (see a).
          The coherent QSNN performs best on the test set (see b).
          And the main advantage of the coherent QSNN is that it can recognize verse1 (see c) and verse2 (see d) with relatively high accuracy.
          Meanwhile, in recognizing 2 normal sentences (see e and f), the QSNNs win only a little since the recognition accuracy of all three kinds of networks is already close to 1. The error bars are drawn as the sample variances. }\label{fig:verse2}
        \end{figure}

        In order to exclude the possibility that the particularity of the training set affects the results displayed in the main text, we provide the training and test results (see Fig.\,\ref{fig:verse2}) of the QSNNs with a new training set and a new test set, where the samples are listed in TABLE \ref{tab:set2}.
        In this case, there are 10 neurons in the hidden layer.
        We also compute the mean Loss and mean accuracy of the networks trained from 15 random initialization QSNN samples $\vec{\gamma}$ or classical NN samples $\vec{w},\vec{b}\ (\gamma_i,w_i,b_i\in[-1, 1]\ \text{uniformly})$.
        Here, the initial learning rates are chosen as $\eta_0^{\vec{h}}=1$ and $\eta_0^{\vec{\gamma}}=\eta_0^{\vec{w}}=\eta_0^{\vec{b}}=2$.

        The result shown in Fig.~\ref{fig:verse2} is similar to the one shown in Fig.~\ref{fig:verse-recognition} in the main text.
        The QSNNs perform better than the classical NN in both training and test, and the coherent QSNN performs best.
        \begin{table*}[htp]
        \caption{Training \& test set} \label{tab:set}
        \begin{ruledtabular}
            \begin{tabular}{p{3cm}|p{13cm}|p{2cm}} %
            \multirow{12}{*}{Training set} &sequence\footnote{Before word sequences are input into the neural networks, all input words in them are lemmatized to their normal forms, and stop words are removed from the word sequences.} &label\\
            \hline
            &There is a gold sun at dawn&Yes\\
            &I love to stay all day in the sun&Yes\\
            &He went for gold that day& Yes\\
            & He loves gold but has nothing& Yes\\
            & He loves the dawn of a day& Yes\\
            & I love the lovely sun& Yes\\
            & sun gold day & No\\
            &day nothing dawn& No\\
            &gold goes love& No\\
            &stay love go sun& No\\
            &day gold nothing& No\\
            &stay dawn love go& No\\
            \hline
            \multirow{4}{*}{Test set}&sequence$^a$ & type\\
            \hline
            &So dawn goes down to day\footnote{Robert Frost. \emph{Nothing Gold Can Stay}}&verse 1\\
            &Nothing gold can stay$^b$&verse 2\\
            &I love to stay here until the dawn&sentence 1\\
            &I love to go out for love&sentence 2\\
            \hline
            word&\multicolumn{2}{c}{gold, sun, dawn, love, stay, day, go, noth}\\
            \end{tabular}
      \end{ruledtabular}
      \end{table*}

         \begin{table*}[htp]
         \caption{Training \& test set} \label{tab:set2}
         \begin{ruledtabular}
            \begin{tabular}{p{3cm}|p{13cm}|p{2cm}} %
            \multirow{16}{*}{Training set} &sequence\footnote{Before word sequences are input into the neural networks, all input words in them are lemmatized to their normal forms, and stop words are removed from the word sequences.} &label\\
            \hline
            &I stand in the sand&Yes\\
            &The traveler loves the gold sand&Yes\\
            &I love the lovely world& Yes\\
            & It stood in the world& Yes\\
            & I love the gold grain& Yes\\
            & I love gold for a long& Yes\\
            & I see the gold sand & Yes\\
            &There is but one world you will see& Yes\\
            &sand gold one& No\\
            &one long world& No\\
            &world grain love& No\\
            &sand stand gold& No\\
            &grain travel sand& No\\
            &grain world love travel& No\\
            &long travel see & No\\
            &see one love&No\\
            \hline
            \multirow{4}{*}{Test set}&sequence$^a$ & type\\
            \hline
            &To see a world in a grain of sand\footnote{William Blake. \emph{Auguries of innocence}}&verse 1\\
            &And be one traveler, long i stood\footnote{Robert Frost. \emph{The Road Not Taken}}&verse 2\\
            &I love to travel over the world&sentence 1\\
            &The traveler stands in gold sand&sentence 2\\
            \hline
            \multirow{2}{*}{word}&\multicolumn{2}{c}{stand, sand, long, travel, love, gold, world, grain, see, one}\\
        \end{tabular}
       \end{ruledtabular}
        \end{table*}

        \subsection{label noise}\label{appendix:details about noise}
        In Sec. \ref{subsec:label noise}, we investigate the robustness of two QSNNs (i.e. coherent QSNN and decoherent QSNN) against label noise in sentence recognition.
        Here, we provide some details about the training and test described in Sec. \ref{subsec:label noise}.
        The structure of the QSNN and correct training set (listed in TABLE \ref{tab:set}) used here are the same as that in the verse recognition (Sec. \ref{sub:verse-recognition}).
        We train the coherent and decoherent QSNN from 4 uniform initialization samples $\{h_i=0.1, \gamma_j=0.1, 0.3, 0.5, 0.7\}_{i,j=1,1}^{m, n}$, where $m$ ($n$) is the number of parameters in the Hamiltonian (Lindblad operators).
        If the label errors are corrected in the number of the iteration $n_{\text{correct}}$, the learning rate here can be given as $\eta=\eta_0/(1+\frac{n_{\text{iteration}}-n_{\text{correct}}}{R})$ when $n_{\text{iteration}} > n_{\text{correct}}$.
        Here, the initial learning rates are chosen as $\eta_0^{\vec{h}}=0.5$ and $\eta_0^{\vec{\gamma}}=3$, and the errors are corrected in the number of iterations $n_{\text{correct}} = 100$.
        The coherent QSNN is more robust against the label noise as shown in Fig.\,\ref{fig:accelerate fitting} of the main text.

    \section{Number of parameters}\label{appendix:number of parameters}
        In a neural network, the number of parameters to be updated usually affects the convergence speed of the model, and is also positively correlated with the resource consumption of the model.
        So, in this part, we would like to exclude the influence of the number of parameters on our results of different QSNNs performances. We take the sentence recognition task described in Sec. \ref{subsec:accelerate} as an example to show the performance improvement of the coherent QSNN is not due to more parameters.

\begin{figure}[h]
          \centering
          \includegraphics[width=8cm]{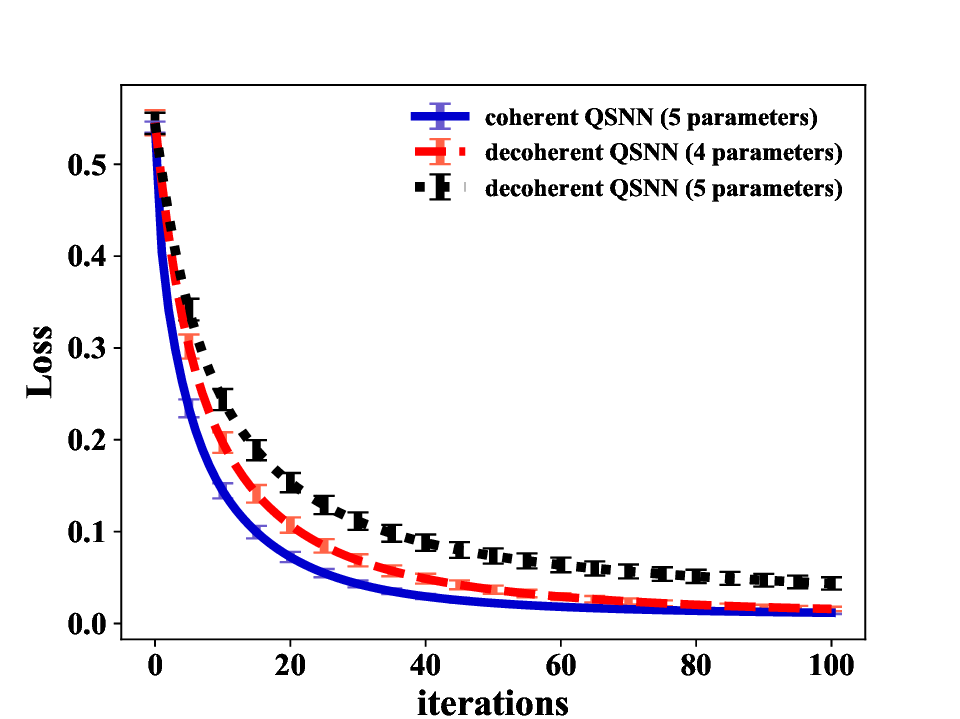}\\
          \caption{A plot of the loss function (the mean of 1000 initialization samples) of the networks with respect to the number of iterations used in the training procedure. Three curves in the figure respectively represent the results of three networks, namely the coherent QSNN with 5 parameters, the decoherent QSNN with 4 parameters, and the decoherent QSNN with 5 parameters. The coherent QSNN requires the fewest training iterations to achieve the same level of performance. The error bars are drawn as 95\% confidence intervals. }\label{fig:paramters_num}
        \end{figure}

        In Sec. \ref{subsec:accelerate}, we have found that the coherent QSNN (5 parameters $\vec{h}\in\mathbb{R}^1, \vec{\gamma}\in\mathbb{R}^4$) requires fewer training iterations than the decoherent one (4 parameters $\gamma\in\mathbb{R}^4$) to achieve the same level of performance. Here, we construct another decoherent QSNN with 5 parameters for comparison. It is constructed by replacing the coherent connection in a coherent QSNN with a decoherent connection. To be specific, we add the Lindblad operators $\{\gamma_5|1\rangle\langle 2|, \gamma_5|2\rangle\langle 1|\}$ to the evolution of the coherent QSNN, at the same time we set $\vec{h}=0$, so that the new decoherent QSNN contains 5 parameters $\vec{\gamma}=(\gamma_1,\gamma_2,...,\gamma_5)$. The training performances of the coherent QSNN and two decoherent QSNNs are shown in Fig. \ref{fig:paramters_num}.
        The coherent QSNN requires the fewest training iterations to achieve the same level of performance, even when compared to the decoherent QSNN with the same number of parameters.
        So, we infer that the training advantage of the coherent QSNN is not a result of more parameters.  It may be due to some physical causes similar to the mechanism of shadow tomography. This relation is worth further exploration.

	\section{Vanishing gradient problem}\label{appendix:vanishing gradient}
        Deep learning based on gradient descent has drawn lots of attention due to its versatile use. However, the vanishing gradient problem is always a stumbling block for its application in both classical and quantum domains. As McClean \emph{et al}. \cite{mcclean2018barren} have demonstrated, the gradient can vanish exponentially in the number of layers in classical deep neural networks \cite{glorot2010understanding,shalev2017failures}, while it may vanish exponentially in the number of qubits in some circuit-model-based quantum neural networks.

        Here, we take the gradient of the loss function with respect to the first dissipation rate $\frac{\partial \text{Loss}}{\partial \gamma_1}$ as an example to investigate the vanishing gradient problem of our model in sentence recognition described in Sec. \ref{sec:quantum advantages}.
        For our QSNN, the gradient $\frac{\partial \text{Loss}}{\partial \gamma_1}$ only depends on the hidden layer number and sentence length.
        The negative effect of the hidden layer number on the gradient can't be avoided in our model as in the classical case.
        However, we are more concerned about the effect of the neuron number in the hidden layer and the sequence length on the gradient here.
        Because these two factors influence whether the QSNN can be used to recognize more complex sentences or not.
         \begin{figure}[ht]
              \centering
              \includegraphics[width=8.5cm]{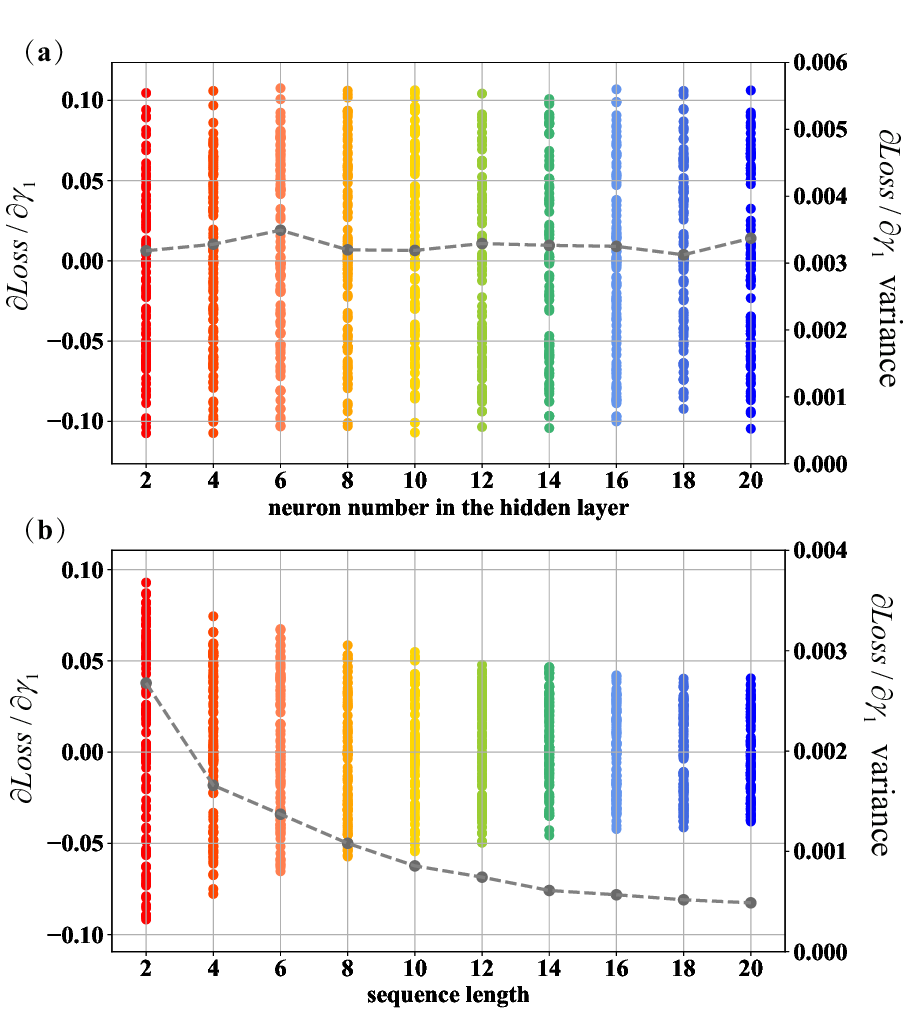}\\
              \caption{The colored dots are drawn by $\frac{\partial \text{Loss}}{\partial \gamma_1}$ computed for 1000 random $\vec{\gamma}$ samples and we choose 100 of them to show. (a) We compute $\frac{\delta \text{Loss}}{\delta \gamma_1}$ of the QSNNs with 1 hidden layer consisting different numbers of neurons. Here, we set $\gamma=0.2,\  T^I=200,\  T^U=T^D=20$. The gray dot represents the sample variance of $\frac{\partial \text{Loss}}{\partial \gamma_1}$ which almost does not change with the number of neurons in the hidden layer. It means there are no barren plateaus in the training landscapes of the QSNNs. (b) In this case, there are 20 neurons in the single hidden layer, and we input sequences labeled Yes with different lengths each time. We show that the gradient decreases with the sequence length. Here, we set $\gamma=0.15,\  T^I=200,\  T^U=T^D=20$.}
              \label{fig:grad-vanishing}
         \end{figure}

        We compute the gradient $\frac{\partial \text{Loss}}{\partial \gamma_1}$ for 1000 random initialization samples $\vec{\gamma}\ (\gamma_i\in[-1, 1]\ \text{uniformly})$ in each case, and uniformly choose 100 results (colored dots of the same color) of them to show in Fig.~\ref{fig:grad-vanishing}.
        The grey dots in Fig.~\ref{fig:grad-vanishing} represent the sample variances of the gradient $\frac{\partial \text{Loss}}{\partial \gamma_1}$.

        First, we investigate the effect of the neuron number in the hidden layer on the gradient $\frac{\partial \text{Loss}}{\partial \gamma_1}$.
        The QSNN we used here only contains one hidden layer with different numbers of hidden layer neurons.
        The training set here only contains the input sequence $(w_1, w_2)$ labeled $\text{Yes}$.
        As shown in Fig.~\ref{fig:grad-vanishing}(a), the sample variance of $\frac{\partial \text{Loss}}{\partial \gamma_1}$ almost does not change with the number of neurons in the hidden layer.
        The number of hidden layer neurons of a QSNN is related to the dimension of the Hilbert space where the state of the QSNN is located.
        So, the gradient doesn't vanish exponentially in the dimension of the Hilbert space, which is different from some circuit-model-based quantum neural networks \cite{mcclean2018barren}.
        From this perspective, the vanishing gradient problem is more tolerable by our QSNN.
        It is because the loss function here is related to the expected value of a global measurement on the final state rather than the expected value of a local Hamiltonian, so Levys lemma-type argumentation \cite{venuti2013probability} does not directly apply.

        Second, we investigate the effect of the sequence length on the gradient $\frac{\partial \text{Loss}}{\partial \gamma_1}$.
        In this case, the QSNN contains one hidden layer with 20 hidden layer neurons.
        We input word sequences of different lengths into the QSNN and calculate the gradient $\frac{\partial \text{Loss}}{\partial \gamma_1}$ for each input. As shown in Fig.~\ref{fig:grad-vanishing}(b), the gradient decreases with the sequence length. However, we think it is not disturbing.
        Because, generally, the lengths of the vast majority of sentences are acceptable, and there is no significant difference in the average length in different corpora.
        So, if we try to train the QSNN with a bigger corpus by adding neurons to the first hidden layer, the gradient won't vanish.


\begin{thebibliography}{53}

\bibitem{lecun2015deep}
Y. LeCun, Y. Bengio, and G. Hinton,
Deep learning,
\href{https://doi.org/10.1038/nature14539}{Nature (London) \textbf{521},~436--444 (2015).}

\bibitem{jordan2015machine}
M. I. Jordan and T. M. Mitchell,
Machine learning: Trends, perspectives, and prospects,
\href{https://doi.org/10.1126/science.aaa8415}{Science \textbf{349}, 255--260 (2015).}

\bibitem{goodfellow2016deep}
I. Goodfellow, Y. Bengio, and A. Courville,
Deep learning,
\href{https://doi.org/10.4258/hir.2016.22.4.351}{(MIT press, 2016).}

\bibitem{nielsen2015neural}
M. A. Nielsen,
Neural networks and deep learning,
\href{https://d1wqtxts1xzle7.cloudfront.net/62971418/neuralnetworksanddeeplearning20200415-115041-1t7vxpc-with-cover-page-v2.pdf?Expires=1631267631&Signature=Ny8D0ZPUFMC-zM1N4G2t5qhOOLmnqpYkC1MYg3nP95OJRLzl2RrsFZciq4h41y2tCt5r2pPS3JKuBhOz7P0Jt5Op-sMP8~shi0Ft7hMZVD8lAKaUdLhrF8XvayK1sOZyaD4a6552W7Zp2nEsAmjwjnNHkxoZYsjKYDecNcP8DXSuxBoZT5yPj1vLmToIp5BuzNWeqjTFy253ieoh6ECLhgickyFTS523d344S4osTHjHClP0Mpo9bn52dM1a2n1r0~fk54-2WJKbC2C~aAGECeTRl8mGjDsZ1Cs5JZnzO9dvB853pe6Bzq20BrZZMUJ74nTlDNi5V9GjCLwvIfV1DA__&Key-Pair-Id=APKAJLOHF5GGSLRBV4ZA}
{(Determination press 2015).}


\bibitem{lecun1998gradient}
Y. Lecun, L. Bottou, Y. Bengio, and P. Haffner,
Gradient-based learning applied to document recognition,
\href{https://doi.org/10.1109/5.726791}{Proceedings of the IEEE \textbf{86}, 2278--2324 (1998).}


\bibitem{le2013building}
Q. V. Le,
Building high-level features using large scale unsupervised learning,
\href{https://doi.org/10.1109/ICASSP.2013.6639343}{IEEE ICASSP 8595--8598, (2013).}

\bibitem{hochreiter1997long}
S. Hochreiter and J. Schmidhuber,
Long short-term memory,
\href{https://doi.org/10.1162/neco.1997.9.8.1735}{Neural computation \textbf{9}, 1735--1780 (1997).}


\bibitem{bengio2003neural}
Y. Bengio, R. Ducharme, P. Vincent, and C. Janvin,
A neural probabilistic language model,
The Journal of Machine Learning Research, \textbf{3}, 1137--1155 (2003).



\bibitem{mikolov2011extensions}
T. Mikolov, S. Kombrink, L. Burget, J. {\v{C}}ernock{\`y}, and S. Khudanpur,
Extensions of recurrent neural network language model,
\href{https://doi.org/10.1109/ICASSP.2011.5947611}{IEEE ICASSP 5528--5531 (2011).}

\bibitem{graves2013speech}
A. Graves, A.-r. Mohamed, and G. Hinton,
Speech recognition with deep recurrent neural networks,
\href{https://doi.org/10.1109/ICASSP.2013.6638947}{IEEE ICASSP 6645--6649 (2013).}

\bibitem{min2017deep}
S. Min, B. Lee, and S. Yoon,
Deep learning in bioinformatics,
\href{https://doi.org/10.1093/bib/bbw068}{Briefings in bioinformatics \textbf{18},  851--869 (2017).}


\bibitem{nielsen2002quantum}
M. A. Nielsen and I. Chuang,
Quantum computation and quantum information,
\href{https://doi.org/10.1119/1.1463744}{(2002).}

\bibitem{ladd2010quantum}
T. D. Ladd, F. Jelezko, R. Laflamme, Y. Nakamura, C. Monroe, and J. L. OBrien,
Quantum computers,
\href{https://doi.org/10.1038/nature08812}{Nature \textbf{464, } 45--53 (2010).}

\bibitem[{Arute et~al.(2019)Arute, Arya, Babbush, Bacon, Bardin, Barends,
  Biswas, Boixo, Brandao, Buell et~al.}]{arute2019quantum}
F. Arute \emph{et~al.},
Quantum supremacy using a programmable superconducting processor,
\href{https://doi.org/10.1038/s41586-019-1666-5}{Nature \textbf{574,} 505--510 ({2019}).}


\bibitem[{Zhong et~al.(2020)Zhong, Wang, Deng, Chen, Peng, Luo, Qin, Wu, Ding,
  Hu et~al.}]{zhong2020quantum}
H.-S. Zhong \emph{et~al.}, Quantum computational advantage using photons,
\href{https://doi.org/10.1126/science.abe8770}{Science \textbf{370,} 1460--1463 (\bibinfo{year}{2020}).}


\bibitem[{Shor(1996)}]{shor1996fault}
P.W. Shor,
\bibinfo{title}{Fault-tolerant quantum computation}, in: \emph{Proceedings of 37th Conference on Foundations of Computer Science},
\bibinfo{pages}{56--65} \bibinfo{year}{}.
\url{https://doi.org/10.1109/SFCS.1996.548464}

\bibitem[{Grover(1997)}]{grover1997quantum}
\bibinfo{author}{L. K. Grover,} \bibinfo{title}{Quantum Mechanics Helps in
  Searching for a Needle in a Haystack},
  \href{https://doi.org/10.1103/PhysRevLett.79.325}{Phys.~Rev.~Lett. \textbf{79}, 325 (\bibinfo{year}{1997}).}

\bibitem[{Harrow et~al.(2009)Harrow, Hassidim, and Lloyd}]{harrow2009quantum}
A. W. Harrow, A. Hassidim, and S. Lloyd,
\bibinfo{title}{Quantum Algorithm for Linear
  Systems of Equations},
\href{https://doi.org/10.1103/PhysRevLett.103.150502}{Phys.~Rev.~Lett. \textbf{103}, 150502 (\bibinfo{year}{2009}).}


\bibitem{montanaro2016quantum}
A Montanaro,
Quantum algorithms: an overview,
\href{https://doi.org/10.1038/npjqi.2015.23}{npj Quantum Inf. \textbf{2}, 15023, (2016).}


\bibitem[{Biamonte et~al.(2017)Biamonte, Wittek, Pancotti, Rebentrost, Wiebe,
  and Lloyd}]{biamonte2017quantum}
J. Biamonte, P. Wittek, N. Pancotti, P. Rebentrost, N. Wiebe, and S. Lloyd,
  \bibinfo{title}{Quantum machine learning},
  \href{https://doi.org/10.1038/nature23474}{Nature \textbf{549,} 195--202 (\bibinfo{year}{2017}).}


\bibitem[{Dunjko and Briegel(2018)}]{dunjko2018machine}
V. Dunjko and H. J. Briegel,
  \bibinfo{title}{Machine learning \& artificial intelligence in the quantum
  domain: a review of recent progress}, \emph{Reports on Progress
  in Physics}, \textbf{81,} 074001 (\bibinfo{year}{2018}).

\bibitem{situ2020quantum}
H.~Situ, Z. He, Y. Wang, L. Li, and S. Zheng,
\bibinfo{title}{Quantum generative adversarial network for generating discrete distribution},
\href{https://doi.org/10.1016/j.ins.2020.05.127}{Information Sciences \textbf{538}, 193--208 (\bibinfo{year}{2020}).}

\bibitem[{Narayanan and Menneer(2000)}]{narayanan2000quantum}
A. Narayanan and T. Menneer,
  \bibinfo{title}{Quantum artificial neural network architectures and
  components},
  \href{https://doi.org/10.1016/S0020-0255(00)00055-4}{Information Sciences \textbf{128}, 231--255 (\bibinfo{year}{2000}).}


\bibitem[{da~Silva et~al.(2016)da~Silva, Ludermir, and
  de~Oliveira}]{da2016quantum}
\bibinfo{author}{A. J. da Silva}, \bibinfo{author}{T. B. Ludermir}, and
  \bibinfo{author}{W. R. de Oliveira}, \bibinfo{title}{Quantum perceptron over
  a field and neural network architecture selection in a quantum computer},
  \href{https://doi.org/10.1016/j.neunet.2016.01.002}{Neural Networks \textbf{76}, 55--64
  (\bibinfo{year}{2016}).}

\bibitem[{Li et~al.(2020)Li, Zhou, Xu, Luo, and Hu}]{li2020quantum}
\bibinfo{author}{Y. Li}, \bibinfo{author}{R.-G. Zhou}, \bibinfo{author}{R. Xu},
  \bibinfo{author}{J. Luo}, and \bibinfo{author}{W. Hu}, \bibinfo{title}{A quantum
  deep convolutional neural network for image recognition},
  \emph{Quantum Science and Technology},
  \textbf{5,} 044003 (\bibinfo{year}{2020}).


\bibitem[{Zeng et~al.(2019)Zeng, Wu, Liu, Wang, and Hu}]{zeng2019learning}
J. Zeng, Y. Wu, J.-G. Liu, L. Wang, and J. Hu,
  \bibinfo{title}{Learning and inference on generative adversarial quantum
  circuits},
  \href{https://doi.org/10.1103/PhysRevA.99.052306}{Phys.~Rev.~A \textbf{99}, 052306 (\bibinfo{year}{2019}).}


\bibitem[{Farhi and Neven(2018)}]{farhi2018classification}
E. Farhi and H. Neven,
  \bibinfo{title}{Classification with quantum neural networks on near term
  processors}, Preprint at \url{https://arxiv.org/abs/1802.06002v2} (2018).

\bibitem[{Mitarai et~al.(2018)Mitarai, Negoro, Kitagawa, and
  Fujii}]{mitarai2018quantum}
K. Mitarai, M. Negoro, M. Kitagawa, and K. Fujii,
  \bibinfo{title}{Quantum circuit learning},
\href{https://doi.org/10.1103/PhysRevA.98.032309}{Phys.~Rev.~A \textbf{98}, 032309 (\bibinfo{year}{2018}).}

\bibitem[{Dallaire-Demers and Killoran(2018)}]{dallaire2018quantum}
P.-L. Dallaire-Demers and N. Killoran,
  \bibinfo{title}{Quantum generative adversarial networks},
  \href{https://doi.org/10.1103/PhysRevA.98.012324}{Phys.~Rev.~A \textbf{98}, 012324 (\bibinfo{year}{2018}).}

\bibitem{killoran2019continuous}
N. Killoran, T. R. Bromley, J. M. Arrazola, M. Schuld,
N. Quesada, and S. Lloyd
Continuous-variable quantum neural networks,
\href{https://doi.org/10.1103/PhysRevResearch.1.033063}{Phys.~Rev.~Research \textbf{1}, 033063 (2019).}

\bibitem{grant2018hierarchical}
E. Grant~\emph{et~al}.,
Hierarchical quantum classifiers,
\href{https://doi.org/10.1038/s41534-018-0116-9}{npj Quantum Information \textbf{4}, 65 (2018).}

\bibitem[{Tacchino et~al.(2019)Tacchino, Macchiavello, Gerace, and
  Bajoni}]{tacchino2019artificial}
F. Tacchino, C. Macchiavello, D. Gerace, and D. Bajoni,
 \bibinfo{title}{An artificial neuron implemented on an actual quantum processor},
  \href{https://doi.org/10.1038/s41534-019-0140-4}{npj Quantum Inf.   \textbf{5}, 26 (\bibinfo{year}{2019}).}


\bibitem[{Wan et~al.(2017)Wan, Dahlsten, Kristj{\'a}nsson, Gardner, and
  Kim}]{wan2017quantum}
K. H. Wan, O. Dahlsten, H. Kristj{\'a}nsson, R. Gardner, and M. Kim,
\bibinfo{title}{Quantum generalisation of
  feedforward neural networks},
  \href{ https://doi.org/10.1038/s41534-017-0032-4}{npj Quantum information \textbf{3}, 36 (\bibinfo{year}{2017}).}


\bibitem{zhao2019building}
J. Zhao, Y.-H. Zhang, C.-P. Shao, Y.-C. Wu, G.-C. Guo, and G.-P. Guo,
  \bibinfo{title}{Building quantum neural networks based on a swap test},
  \href{https://doi.org/10.1103/PhysRevA.100.012334}{Phys.~Rev.~A \textbf{100}, 012334 (\bibinfo{year}{2019}).}


\bibitem {he2021variational}
\bibinfo{author}{Z. He}, \bibinfo{author}{L.~Li}, \bibinfo{author}{S.~Zheng}, \bibinfo{author}{Y.~Li}, and  \bibinfo{author}{H.~Situ},
\bibinfo{title}{Variational quantum compiling with double Q-learning},
\emph{New Journal of Physics},
\textbf{23}, 033002 (\bibinfo{year}{2021}).

\bibitem{beer2020training}
\bibinfo{author}{K. Beer} \emph{et~al}., \bibinfo{title}{Training deep quantum neural
  networks},
  \href{https://doi.org/10.1038/s41467-020-14454-2}{Nat. commun. \textbf{11}, 808 (\bibinfo{year}{2020}).}

\bibitem{steinbrecher2019quantum}
G. R. Steinbrecher, J. P. Olson, D. Englund, and J. Carolan,
  \bibinfo{title}{Quantum optical neural networks},
  \href{https://doi.org/10.1038/s41534-019-0174-7}{npj
  Quantum Inf. \textbf{5}, 60
  (\bibinfo{year}{2019}).}

\bibitem{cong2019quantum}
I. Cong, S. Choi, and M. D. Lukin,
 \bibinfo{title}{Quantum convolutional neural networks},
\href{https://doi.org/10.1038/s41567-019-0648-8}{Nat. Phys. \textbf{15}, 1273--1278 (\bibinfo{year}{2019}).}


\bibitem[{Dalla~Pozza and Caruso(2020)}]{dalla2020quantum}
N. Dalla Pozza and F. Caruso,
  \bibinfo{title}{Quantum state discrimination on reconfigurable noise-robust
  quantum networks},
  \href{https://doi.org/10.1103/PhysRevResearch.2.043011}{Phys.~Rev.~Research \textbf{2}, 043011 (\bibinfo{year}{2020}).}


\bibitem{schuld2014quantum}
 M. Schuld, I. Sinayskiy, and F. Petruccione,
 \bibinfo{title}{Quantum walks on graphs
  representing the firing patterns of a quantum neural network},
  \href{https://doi.org/10.1103/PhysRevA.89.032333}{Phys.~Rev.~A \textbf{89}, 032333 (\bibinfo{year}{2014}).}


\bibitem[{Rebentrost et~al.(2018)Rebentrost, Bromley, Weedbrook, and
  Lloyd}]{rebentrost2018quantum}
P. Rebentrost, T. R. Bromley, C. Weedbrook, and S. Lloyd,
  \bibinfo{title}{Quantum Hopfield neural network}.
  \href{https://doi.org/10.1103/PhysRevA.98.042308}{Phys.~Rev.~A \textbf{98}, 042308 (\bibinfo{year}{2018}).}

\bibitem{amin2018quantum}
M. H. Amin, E. Andriyash, J. Rolfe, B. Kulchytskyy, and R. Melko,
Quantum Boltzmann Machine,
\href{https://doi.org/10.1103/PhysRevX.8.021050}{Phys.~Rev.~X \textbf{8}, 021050 (2018).}


\bibitem[{Tang et~al.(2019)Tang, Feng, Wang, Lai, Wang, Ye, Wang, Shi, Wang,
  Chen et~al.}]{tang2019experimental}
H. Tang, Z. Feng, Y.-H. Wang, P.-C. Lai, C.-Y. Wang, Z.-Y. Ye, C.-K. Wang, Z.-Y. Shi, T.-Y. Wang, Y. Chen, J. Gao, and X.-M. Jin,
\bibinfo{title}{Experimental quantum
  stochastic walks simulating associative memory of Hopfield neural networks},
  \href{https://doi.org/10.1103/PhysRevApplied.11.024020}{Phys.~Rev.~Applied \textbf{11}, 024020 (\bibinfo{year}{2019}).}


\bibitem[{Carleo and Troyer(2017)}]{carleo2017solving}
G. Carleo1 and M. Troyer,
  \bibinfo{title}{Solving the quantum many-body problem with artificial neural
  networks},
  \href{https://doi.org/10.1126/science.aag2302}{Science \textbf{355}, 602--606 (\bibinfo{year}{2017}).}

\bibitem[{Gao and Duan(2017)}]{gao2017efficient}
 X. Gao and L.-M. Duan,
  \bibinfo{title}{Efficient representation of quantum many-body states with
  deep neural networks},
  \href{https://doi.org/10.1038/s41467-017-00705-2}{Nat. commun. \textbf{8,} 662 (\bibinfo{year}{2017}).}


\bibitem[{Bondarenko and Feldmann(2020)}]{bondarenko2020quantum}
D. Bondarenko and P. Feldmann,
  \bibinfo{title}{Quantum Autoencoders to Denoise Quantum Data}.
  \href{https://doi.org/10.1103/PhysRevLett.124.130502}{Phys.~Rev.~Lett. \textbf{124,} 130502 (\bibinfo{year}{2020}).}


\bibitem[{Zhang et~al.(2018{\natexlab{b}})Zhang, Niu, Su, Wang, Ma, and
 Song}]{zhang2018end}
P. Zhang, J. Niu, Z. Su, B. Wang, L. Ma, and D. Song,
\bibinfo{title}{End-to-end quantum-like language models with application to question answering}, in:
  \emph{Proceedings of the AAAI Conference on Artificial
  Intelligence} \textbf{32,} \bibinfo{year}{(2018)}.

\bibitem[{Sordoni et~al.(2013)Sordoni, Nie, and Bengio}]{sordoni2013modeling}
A. Sordoni, J.-Y. Nie, and Y. Bengio,
 \bibinfo{title}{Modeling term dependencies with
  quantum language models for ir}, in: \emph{Proceedings of the
  36th international ACM SIGIR conference on Research and development in
  information retrieval}, \bibinfo{pages}{653--662} \bibinfo{year}{(2013)}.
  \url{https://doi.org/10.1145/2484028.2484098}

\bibitem[{Basile and Tamburini(2017)}]{basile2017towards}
 I. Basile and F. Tamburini,
  \bibinfo{title}{Towards quantum language models}, in:
  \emph{Proceedings of the 2017 Conference on Empirical Methods
  in Natural Language Processing}, \bibinfo{pages}{1840--1849} \bibinfo{year}{(2017)}.
  \url{https://doi.org/10.18653/v1/D17-1196}

\bibitem[{Zhang et~al.(2019)Zhang, Song, Zhang, Li, and
  Wang}]{zhang2019quantum}
 Y. Zhang, D. Song, P. Zhang, X. Li, and P. Wang,
 \bibinfo{title}{A quantum-inspired sentiment representation model for twitter sentiment analysis}.
  \href{https://doi.org/10.1007/s10489-019-01441-4}{Applied Intelligence \textbf{49}, 3093--3108 (\bibinfo{year}{2019}).}

\bibitem[{Whitfield, James D and Rodr{\'\i}guez-Rosario, C{\'e}sar A and Aspuru-Guzik, Al{\'a}n}]{whitfield2010quantum}
J. D. Whitfield, C. A. Rodr{\'\i}guez-Rosario, and A. Aspuru-Guzik,
\bibinfo{title}{Quantum stochastic walks: A generalization of classical random walks and quantum walks},
\href{https://doi.org/10.1103/PhysRevA.81.022323}{Phys. Rev. A \textbf{81}, 022323 (\bibinfo{year}{2010}).}


\bibitem{kossakowski1972quantum}
A. Kossakowski,
\bibinfo{title}{On quantum statistical mechanics of non-Hamiltonian systems},
\href{https://doi.org/10.1016/0034-4877(72)90010-9}{Rep. Math. Phys. \textbf{3}, 247 (1972).}

\bibitem[{Lindblad(1976)}]{lindblad1976generators}
\bibinfo{author}{G. Lindblad}, \bibinfo{title}{On the generators of quantum
  dynamical semigroups},
  \href{https://doi.org/10.1007/BF01608499}{Commun. Math. Phys. \textbf{48}, 119--130 (\bibinfo{year}{1976}).}

\bibitem{gorini1976completely}
\bibinfo{author}{V. Gorini, A. Kossakowski, and E. C. G. Sudarshan},
Completely positive dynamical semigroups of N-level systems,
\href{https://doi.org/10.1063/1.522979}{J. Math. Phys. \textbf{17}, 821 (1976).}

\bibitem{cybenko1989approximation}
G. Cybenko,
Approximation by superpositions of a sigmoidal function,
\emph{Mathematics of control, signals and systems}, \textbf{2}, 303--314, (1989).


\bibitem{hornik1989multilayer}
K. Hornik, M. Stinchcombe, and H. White,
Multilayer feedforward networks are universal approximators,
\href{https://doi.org/10.1016/0893-6080(89)90020-8}{Neural networks \textbf{2}, 359--366, (1989).}

\bibitem{zhang2000neural}
G.P. Zhang,
Neural networks for classification: a survey,
\emph{IEEE Transactions on Systems, Man, and Cybernetics, Part C (Applications and Reviews)}, \textbf{30}, 451--462, (2000).
\url{https://doi.org/10.1109/5326.897072}

\bibitem{livni2014computational}
R. Livni,  S. Shalev-Shwartz, and O. Shamir,
On the computational efficiency of training neural networks,
in \emph{Proceedings of the 27th International Conference on Neural Information Processing Systems}, Vol.1 (MIT Press, Cambridge, MA), pp. 855-863.

\bibitem{aaronson2019shadow}
S. Aaronson,
Shadow tomography of quantum states,
\href{https://dl.acm.org/doi/abs/10.1145/3188745.3188802}{SIAM Journal on Computing \textbf{49}, STOC18--368, (2020).}

\bibitem{kunjummen2021shadow}
J. Kunjummen, M. C. Tran, D. Carney, and J. M. Taylor,
Shadow process tomography of quantum channels,
Preprint at \url{https://arxiv.org/pdf/2110.03629.pdf (2021).}

\bibitem{huang2021information}
H. Y. Huang, R. Kueng, and J. Preskill,
Information-Theoretic Bounds on Quantum Advantage in Machine Learning,
\href{https://journals.aps.org/prl/abstract/10.1103/PhysRevLett.126.190505}{Phys. Rev. Lett. \textbf{126}, 190505, (2021).}

\bibitem{huang2021provably}
H. Y. Huang, R. Kueng, G. Torlai, V. V. Albert, and J. Preskill,
Provably efficient machine learning for quantum many-body problems,
Preprint at \url{https://arxiv.org/pdf/2106.12627.pdf (2021).}


\bibitem{huang2021quantum}
H. Y. Huang, M. Broughton, J. Cotler, S. Chen, J. Li, M. Mohseni, H. Neven, R. Babbush, R. Kueng, J. Preskill, and J. R. McClean,
Quantum advantage in learning from experiments,
Preprint at \url{https://arxiv.org/pdf/2112.00778.pdf (2021).}


\bibitem[{Glorot and Bengio(2010)}]{glorot2010understanding}
X. Glorot and Y. Bengio,
  \bibinfo{title}{Understanding the difficulty of training deep feedforward
  neural networks}, in \emph{Proceedings of the thirteenth
  international conference on artificial intelligence and statistics} (ML Research Press, 2010),
pp. 249--256.

\bibitem[{Shalev-Shwartz et~al.(2017)Shalev-Shwartz, Shamir, and
  Shammah}]{shalev2017failures}
S. Shalev-Shwartz, O. Shamir, and S. Shammah, \bibinfo{title}{Failures of gradient-based deep learning},
  in \emph{Proceedings of the 34th International Conference on Machine Learning (International Convention Centre, Sydney, Australia, 2017)}, Vol. 70 of Proceedings of Machine Learning Research (PMLR, 2017), pp.3067–3075.

\bibitem[{McClean et~al.(2018)McClean, Boixo, Smelyanskiy, Babbush, and
  Neven}]{mcclean2018barren}
J. R. McClean, S. Boixo, V. N. Smelyanskiy, R. Babbush, and H. Neven,
  \bibinfo{title}{Barren plateaus in quantum neural network training landscapes},
  \href{https://doi.org/10.1038/s41467-018-07090-4}{Nat. commun. \textbf{9}, 4812 (\bibinfo{year}{2018}).}

\bibitem[{Du et~al.(2003)Du, Li, Xu, Shi, Wu, Zhou, and
  Han}]{du2003experimental}
J. Du, H. Li, X. Xu, M. Shi, J. Wu, X. Zhou, and R. Han,
\bibinfo{title}{Experimental implementation of the quantum random-walk algorithm},
  \href{https://doi.org/10.1103/PhysRevA.67.042316}{Phys.~Rev.~A \textbf{67}, 042316 (\bibinfo{year}{2003}).}


\bibitem[{Tang et~al.(2018)Tang, Lin, Feng, Chen, Gao, Sun, Wang, Lai, Xu, Wang
  et~al.}]{tang2018experimental}
\bibinfo{author}{H Tang} \emph{et~al}., \bibinfo{title}{Experimental
  two-dimensional quantum walk on a photonic chip},
  \href{https://doi.org/10.1126/sciadv.aat3174}{Science advances \textbf{4}, eaat3174 (\bibinfo{year}{2018}).}


\bibitem[{Perets et~al.(2008)Perets, Lahini, Pozzi, Sorel, Morandotti, and
  Silberberg}]{perets2008realization}
H. B. Perets, Y. Lahini, F. Pozzi, M. Sorel, R. Morandotti, and Y. Silberberg,
\bibinfo{title}{Realization of Quantum Walks with Negligible Decoherence in Waveguide Lattices},
  \href{https://doi.org/10.1103/PhysRevLett.100.170506}{Phys.~Rev.~Lett. \textbf{100}, 170506 (\bibinfo{year}{2008}).}

\bibitem{laneve2021experimental}
A. Laneve, A. Geraldi, F. Hamiti, P. Mataloni, and F. Caruso,
Experimental multi-state quantum discrimination through a Quantum network,
Preprint at \url{https://arxiv.org/abs/2107.09968 (2021).}

\bibitem{venuti2013probability}
L. C. Venuti and P. Zanardi,
Probability density of quantum expectation values,
\href{https://doi.org/10.1016/j.physleta.2013.05.041}{Physics Letters A \textbf{377}, 1854--1861 (2013).}


\end{thebibliography}
\end{document}